\documentclass[a4paper,11pt]{article}
\pdfoutput=1
\usepackage{jheppub}
\usepackage[utf8]{inputenc}
\usepackage{amsmath}
\usepackage{amsfonts}
\usepackage{amssymb}
\usepackage{physics}
\usepackage{tikz}
\usepackage{appendix}
\usepackage{hyperref}
\usepackage{xcolor}
\usepackage{graphicx}
\usepackage{float}

\newcommand{\fft}[2]{\frac{#1}{#2}}
\newcommand{\ft}[2]{{\textstyle{\frac{#1}{#2}}}}
\newcommand{\nn}{\nonumber}
\newcommand{\Ai}{\operatorname{Ai}}
\renewcommand{\Re}{\operatorname{Re}}
\renewcommand{\Im}{\operatorname{Im}}
\newcommand{\Li}{\operatorname{Li}}
\newcommand{\Cl}{\operatorname{Cl}}
\newcommand{\ol}[1]{\overline{#1}}
\newcommand{\cN}{\mathcal{N}}

\newcommand{\cO}{\mathcal{O}}

\numberwithin{equation}{section}

\begin{document}
\preprint{LCTP-21-12}

\title{Subleading Corrections in $\cN=3$ Gaiotto-Tomasiello Theory}
\author{James T. Liu and Robert J. Saskowski}

\affiliation{Leinweber Center for Theoretical Physics, Randall Laboratory of Physics\\The University of Michigan, Ann Arbor, MI 48109-1040, USA }

\emailAdd{jimliu@umich.edu, rsaskows@umich.edu}

\abstract{
    We study subleading corrections to the genus-zero free energy of the $\mathcal{N}=3$ Gaiotto-Tomasiello theory. In general, we obtain the endpoints and free energy as a set of parametric equations via contour integrals of the planar resolvent, up to exponentially suppressed corrections. In the particular case that the two gauge groups in the quiver are of equal rank, we find an explicit (perturbative) expansion for the free energy. If, additionally, both groups have equal levels, then we find the full expression for the genus-zero free energy, modulo exponentially suppressed corrections. We also verify our results numerically.
}

\maketitle

\section{Introduction}

The AdS/CFT correspondence conjectures a remarkable equivalence between large-$N$ gauge theories and string/M-theory on asymptotically AdS backgrounds.  In this context, Chern-Simons-matter theories are of particular interest in regards to the dynamics of M2-branes \cite{Bagger:2006sk,Bagger:2007jr,Bagger:2007vi,Gustavsson:2007vu,Gustavsson:2008dy,VanRaamsdonk:2008ft,Aharony:2008ug}. In particular, the worldvolume theory of $N$ coincident M2-branes probing the singularity of a $\mathbb{C}^4/\mathbb{Z}_k$ orbifold was constructed in \cite{Aharony:2008ug} and is known as the Aharony-Bergman-Jafferis-Maldacena (ABJM) theory. ABJM theory is an $\mathcal N=6$, $U(N)_k\times U(N)_{-k}$ Chern-Simons-matter theory, and in the large-$N$ limit is dual to either M-theory on AdS$_4\times S^7/\mathbb{Z}_k$ or IIA string theory on AdS$_4\times\mathbb{CP}^3$, depending on the limit taken.

ABJM theory and its holographic dual provide an excellent opportunity to probe the dynamics of string/M-theory as well as quantum gravity and AdS$_4$ black holes.  However, as AdS/CFT is a strong/weak coupling duality, it is highly non-trivial to make direct comparisons on both sides of the duality.  Nevertheless, certain path integrals in superconformal Chern-Simons-matter theories reduce to matrix models via supersymmetric localization \cite{Pestun:2007rz,Kapustin:2009kz}. Such localization techniques have long been studied in the context of supersymmetric and topological QFTs, and the application of \cite{Pestun:2007rz,Kapustin:2009kz} to superconformal field theories have proven a powerful technique to analyze observables via matrix models. In particular, ABJM theory can be localized to a two-matrix model \cite{Kapustin:2009kz}, which can then be studied via standard methods of random matrix theory or by novel methods such as the ideal Fermi gas approach \cite{Marino:2011eh}.

Many important results have been obtained for the supersymmetric partition function and Wilson loop observables in ABJM theory \cite{Marino:2009jd,Marino:2011eh,Hatsuda:2012hm,Putrov:2012zi} and the ABJ generalization \cite{Awata:2012jb,Honda:2014npa,Hatsuda:2016rmv,Cavaglia:2016ide}.  In particular, the $S^3$ partition function at fixed Chern-Simons levels $k$ and $-k$ was shown to have the form of an Airy function
\begin{equation}
    Z_{\mathrm{ABJM}}^{S^3}=\left(\fft2{\pi^2k}\right)^{-1/3}e^{A(k)}\Ai\left[\left(\fft2{\pi^2k}\right)^{-1/3}\left(N-\fft1{3k}-\fft{k}{24}\right)\right]+Z_{\mathrm{np}},
\end{equation}
where $A(k)$ encodes certain quantum corrections and $Z_{\mathrm{np}}$ is a non-perturbative contribution.  Taking $F=\log Z$ then leads to a fixed $k$ expansion of the free energy as
\begin{equation}
    F_{\mathrm{ABJM}}=\fft{\pi\sqrt2}3k^{1/2}N^{3/2}-\fft\pi{\sqrt{2k}}\left(\fft{k^2}{24}+\fft13\right)N^{1/2}+\fft14\log N+\mathcal O(1).
\end{equation}
In the M-theory dual, the $N^{3/2}$ term can be matched to the on-shell classical supergravity action, while the $N^{1/2}$ term is related to eight-derivative couplings in M-theory \cite{Bergman:2009zh,Aharony:2009fc} which reduce to four-derivative couplings in AdS$_4$ supergravity \cite{Bobev:2020egg,Bobev:2021oku}.  The Airy function form of the partition function holds for a wide range of Chern-Simons-matter theories beyond ABJM theory.  Then, by expanding the Airy function at large $N$, one can see that the $\fft14\log N$ term is universal to this full set of theories.  As an important test of quantum gravity, this log term has been reproduced successfully by a one-loop calculation in eleven-dimensional supergravity on AdS$_4\times X^7$ \cite{Bhattacharyya:2012ye}.

Given the remarkable successes of precision tests of ABJM holography, we wish to extend such investigations to the Gaiotto-Tomasiello (GT) case \cite{Gaiotto:2009mv}.  The GT theory is an $\mathcal{N}=3$ Chern-Simons-matter theory, and can be thought of as a generalization of the ABJM theory to arbitrary Chern-Simons levels, $k_1$ and $k_2$, with $F_0=k_1+k_2\ne0$.  This model is dual to massive IIA supergravity with $F_0$ playing the role of the Romans mass \cite{Romans:1985tz}.  The leading order behavior of GT free energy is \cite{Suyama:2010hr,Suyama:2011yz,Jafferis:2011zi}
\begin{equation}
    F_{\mathrm{GT}}=\fft{3^{5/3}\pi }{5\cdot2^{4/3}}e^{-i\pi/6}(k_1+k_2)^{1/3}N^{5/3}+\cdots.
\label{eq:FGT0}
\end{equation}
The $N^{5/3}k^{1/3}$ scaling is in contrast to the $N^{3/2}k^{1/2}$ scaling of the ABJM free energy, and has confirmed on the supergravity side \cite{Aharony:2010af}.  While this leading-order behavior is well established and generalizes to a large class of $\mathcal N=3$ necklace quiver models with $F_0\ne0$, less is known about its subleading corrections, which is the focus of this paper.

Although the partition function for GT theory can also be mapped to a corresponding ideal Fermi gas system, unlike for the ABJM model, the resulting expression does not take the form of an Airy function \cite{Marino:2011eh,Hong:2021bsb}.  Furthermore, the mapping to the quantum Fermi gas system promoted in \cite{Marino:2011eh} involves taking
\begin{equation}
    \fft{4\pi}\hbar=\fft1{k_1}-\fft1{k_2}.
\end{equation}
This demonstrates that a small $\hbar$ expansion is in tension with taking $k_1\approx k_2$, which is the natural realm for exploring the free energy in (\ref{eq:FGT0}).  We thus find it more natural to work directly with the GT theory partition function written as a two-matrix model.  While a saddle point analysis was performed in \cite{Hong:2021bsb}, here we use a standard resolvent approach and compute the genus-zero partition function as an expansion in inverse powers of the 't~Hooft parameter $t=g_sN$ with $g_s=2\pi i/k$ where $k$ is an effective overall Chern-Simons level.  For equal levels, $k=k_1=k_2$, we find (at genus zero)
\begin{equation}
    F_{\mathrm{GT}}^{k_1=k_2}=\fft1{g_s^2}\left[\fft35\left(\fft{3\pi^2}2\right)^{2/3}\left(t+\fft{\zeta(3)}{2\pi^2}\right)^{5/3}-\fft{\pi^2}{12}t+\mbox{const.}\right],
\label{eq:FGT=}
\end{equation}
up to exponentially small corrections in the large $|t|$ limit.

To highlight the first subleading corrections to the planar free energy, we  substitute $t=2\pi iN/k$ into (\ref{eq:FGT=}) and expand to obtain
\begin{equation}
    F_{\mathrm{GT}}^{k_1=k_2}=\fft{3^{5/3}\pi}{10}e^{-i\pi/6}k^{1/3}N^{5/3}+\fft{i\pi}{24}kN+\fft{3^{2/3}}{8\pi^2}e^{-2\pi i/3}\zeta(3)k^{4/3}N^{2/3}+\mathcal O(1).
\end{equation}
The leading order $N^{5/3}$ term matches (\ref{eq:FGT0}), while the linear-$N$ term was previously obtained in \cite{Hong:2021bsb}, and is pure imaginary for real Chern-Simons levels.  At the next order, we find a $N^{2/3}$ term with a coefficient proportional to $\zeta(3)$.  This term is of $\mathcal O(1/t)$ compared to the leading order, and has a natural interpretation in the massive IIA supergravity dual as originating from a tree-level $\alpha'^3R^4$ coupling.

This paper is organized as follows. In Section~\ref{sec:GTreview}, we predominantly follow \cite{Suyama:2010hr} in summarizing important results about the planar limit and the resolvent in GT theory.  We then proceed in Section~\ref{sec:GTfree} to obtain the planar free energy from the resolvent in the limit of large 't~Hooft coupling, and further check our results against numerical data.  Finally, we conclude in Section~\ref{sec:disc} with some open questions.  Some of the more technical calculations are relegated to two appendices.

\section{GT theory and the planar resolvent}
\label{sec:GTreview}

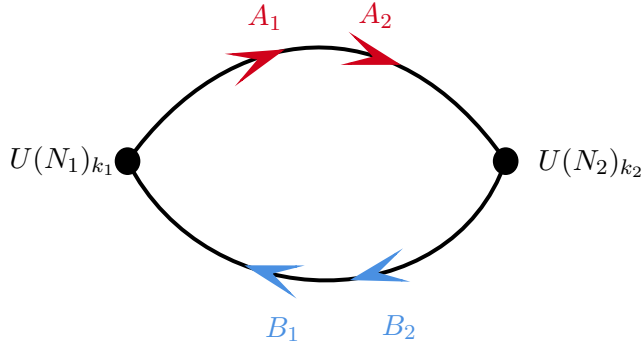
\begin{figure}[t]
    \centering
    \tikzset{every picture/.style={line width=0.75pt}} 
    
    \begin{tikzpicture}[x=0.75pt,y=0.75pt,yscale=-.75,xscale=.75]
    
    \draw  [fill={rgb, 255:red, 0; green, 0; blue, 0 }  ,fill opacity=1 ] (149.51,129.68) .. controls (149.51,124.93) and (153.03,121.07) .. (157.38,121.07) .. controls (161.73,121.07) and (165.25,124.93) .. (165.25,129.68) .. controls (165.25,134.44) and (161.73,138.3) .. (157.38,138.3) .. controls (153.03,138.3) and (149.51,134.44) .. (149.51,129.68) -- cycle ;
    \draw  [fill={rgb, 255:red, 0; green, 0; blue, 0 }  ,fill opacity=1 ] (401.18,129.68) .. controls (401.18,124.93) and (404.7,121.07) .. (409.05,121.07) .. controls (413.4,121.07) and (416.92,124.93) .. (416.92,129.68) .. controls (416.92,134.44) and (413.4,138.3) .. (409.05,138.3) .. controls (404.7,138.3) and (401.18,134.44) .. (401.18,129.68) -- cycle ;
    \draw [color={rgb, 255:red, 0; green, 0; blue, 0 }  ,draw opacity=1 ][line width=1.5]    (157.38,129.68) .. controls (211.53,236.99) and (372.55,235.68) .. (409.05,129.68) ;
    \draw [color={rgb, 255:red, 0; green, 0; blue, 0 }  ,draw opacity=1 ][line width=1.5]    (157.38,129.68) .. controls (224.65,34.65) and (334.38,21.6) .. (409.05,129.68) ;
    \draw  [color={rgb, 255:red, 208; green, 2; blue, 27 }  ,draw opacity=1 ][fill={rgb, 255:red, 208; green, 2; blue, 27 }  ,fill opacity=1 ] (227.29,57.97) -- (259.97,55.51) -- (232.87,75.63) -- (245.03,61.16) -- cycle ;
    \draw  [color={rgb, 255:red, 208; green, 2; blue, 27 }  ,draw opacity=1 ][fill={rgb, 255:red, 208; green, 2; blue, 27 }  ,fill opacity=1 ] (308.8,46.42) -- (337.13,64.42) -- (304.37,64.47) -- (321.86,59.93) -- cycle ;
    \draw  [color={rgb, 255:red, 74; green, 144; blue, 226 }  ,draw opacity=1 ][fill={rgb, 255:red, 74; green, 144; blue, 226 }  ,fill opacity=1 ] (265.61,218.32) -- (237.35,200.19) -- (270.11,200.29) -- (252.6,204.75) -- cycle ;
    \draw  [color={rgb, 255:red, 74; green, 144; blue, 226 }  ,draw opacity=1 ][fill={rgb, 255:red, 74; green, 144; blue, 226 }  ,fill opacity=1 ] (340.36,209.73) -- (307.62,208.85) -- (336.32,191.57) -- (322.98,204.75) -- cycle ;
    
    \draw (235.16,23.26) node [anchor=north west][inner sep=0.75pt]  [color={rgb, 255:red, 208; green, 2; blue, 27 }  ,opacity=1 ]  {$A_{1}$};
    \draw (309.11,21.96) node [anchor=north west][inner sep=0.75pt]  [color={rgb, 255:red, 208; green, 2; blue, 27 }  ,opacity=1 ]  {$A_{2}$};
    \draw (247.09,232.04) node [anchor=north west][inner sep=0.75pt]  [color={rgb, 255:red, 74; green, 144; blue, 226 }  ,opacity=1 ]  {$B_{1}$};
    \draw (324.62,230.74) node [anchor=north west][inner sep=0.75pt]  [color={rgb, 255:red, 74; green, 144; blue, 226 }  ,opacity=1 ]  {$B_{2}$};
    \draw (76.8,117.47) node [anchor=north west][inner sep=0.75pt]    {$U( N_{1})_{k_{1}}$};
    \draw (428.67,118.78) node [anchor=north west][inner sep=0.75pt]    {$U( N_{2})_{k_{2}}$};
    \end{tikzpicture}

    \caption{The $\cN=3$ GT quiver diagram. $A_1$ and $A_2$ are bifundamental hypermultiplets and $B_1$ and $B_2$ are anti-bifundamental hypermultiplets coupling the nodes of the quiver.}
    \label{fig:quiver}
\end{figure}

GT theory is an $\mathcal{N}=3$ superconformal Chern-Simons-matter theory with $U(N_1)_{k_1}\times U(N_2)_{k_2}$ gauge group and quiver diagram given in Figure~\ref{fig:quiver}.  It was originally constructed as a deformation of ABJM theory in \cite{Gaiotto:2009mv} by allowing the two $U(N)$ quivers to take on arbitrary ranks and levels, which in turn knocks the supersymmetry down from $\mathcal{N}=6$ to $\mathcal{N}=3$. On the dual gravity side, which was constructed to first order in perturbation theory in \cite{Gaiotto:2009yz}, this corresponds to turning on a nonzero Romans mass $F_0=k_1+k_2$, which is a 0-form R-R flux sourced by D8-branes.  The supergravity description then corresponds to the massive IIA theory where the 2-form NS-NS $B$-field acquires a mass precisely equal to $F_0$ by ``eating'' the 1-form gauge field in a Higgs-like mechanism \cite{Romans:1985tz}. It is generally believed that there is no M-theory limit \cite{Aharony:2010af} when this mass is non-vanishing.

Since GT theory still retains $\mathcal{N}=3$ supersymmetry, its partition function can be localized following \cite{Kapustin:2009kz}, just as in the AJBM case.  The resulting matrix model takes the form
\begin{equation}
    Z=\fft1{N_1!N_2!}\int\prod_{i=1}^{N_1}\frac{\dd{u_i}}{2\pi}\prod_{j=1}^{N_2}\frac{\dd{v_j}}{2\pi}e^{-S(u_i,v_j)},
\end{equation}
where the action is given by
\begin{equation}
    e^{-S}=\exp\qty[\frac{ik_1}{4\pi}\sum_{i=1}^{N_1}u_i^2+\frac{ik_2}{4\pi}\sum_{i=1}^{N_2}v_i^2]\frac{\prod_{i<j}^{N_1}\sinh^2\qty(\frac{u_i-u_j}{2})\prod_{i<j}^{N_2}\sinh^2\qty(\frac{v_i-v_j}{2})}{\prod_{i=1}^{N_1}\prod_{j=1}^{N_2}\cosh^2\qty(\frac{u_i-v_j}{2})}.
\end{equation}
Since there are two independent Chern-Simons levels, $k_1$ and $k_2$, we can define two 't~Hooft couplings, $\lambda_1=N_1/k_1$ and $\lambda_2=N_2/k_2$.  However, to highlight the planar limit, we find it more convenient to follow \cite{Suyama:2010hr} by introducing an auxiliary parameter $k$ and defining
\begin{equation}
    t_1=\fft{2\pi iN_1}k,\qquad t_2=\fft{2\pi iN_2}k,\qquad\kappa_1=\fft{k_1}k,\qquad\kappa_2=\fft{k_2}k.
\end{equation}
The planar limit is then taken by sending $k\to\infty$ while holding $t_i$ and $\kappa_i$ fixed.

Written in terms of the above quantities, the action now takes the form%
\footnote{Note that this choice of parameters differs from that of \cite{Suyama:2010hr} in the choice of sign of $t_2$ and $\kappa_2$.  In particular, $(t_2)_{\mathrm{there}}=(-t_2)_{\mathrm{here}}$ and $(\kappa_2)_{\mathrm{there}}=(-\kappa_2)_{\mathrm{here}}$.}
\begin{align}
    S&=\fft1{g_s^2}\Bigl[\fft{\kappa_1t_1}{2N_1}\sum_{i=1}^{N_1}u_i^2+\fft{\kappa_2t_2}{2N_2}\sum_{i=1}^{N_2}v_i^2-\fft{t_1^2}{N_1^2}\sum_{i<j}^{N_1}\log\sinh^2\fft{u_i-u_j}2-\fft{t_2^2}{N_2^2}\sum_{i<j}^{N_2}\log\sinh^2\fft{v_i-v_j}2\nn\\
    &\kern14em+\fft{t_1t_2}{N_1N_2}\sum_{i=1}^{N_1}\sum_{j=1}^{N_2}\log\cosh^2\fft{u_i-v_j}2\Bigr],
\label{eq:action}
\end{align}
where we have introduced $g_s=2\pi i/k$.  While the physical Chern-Simons levels $k_1$ and $k_2$ are real, below we will analytically continue to imaginary levels such that the couplings $t_i$ and $\kappa_i$ are real.  This will allow us to work with a real action and corresponding real saddle point equations.  In particular, varying the action, (\ref{eq:action}), with respect to $u_i$ and $v_j$ gives the saddle-point equations
\begin{subequations}
\begin{align}
     \kappa_{1} u_{i} &=\frac{t_{1}}{N_{1}} \sum_{j \neq i}^{N_{1}} \operatorname{coth} \frac{u_{i}-u_{j}}{2}-\frac{t_{2}}{N_{2}} \sum_{j=1}^{N_{2}} \tanh \frac{u_{i}-v_{j}}{2}, \\
    \kappa_{2} v_{i} &=\frac{t_{2}}{N_{2}} \sum_{j \neq i}^{N_{2}} \operatorname{coth} \frac{v_{i}-v_{j}}{2}-\frac{t_{1}}{N_{1}} \sum_{j=1}^{N_{1}} \tanh \frac{v_{i}-u_{j}}{2}.
\end{align}
\label{eq:unexponentatedSPE}
\end{subequations}
At this stage, it is convenient to switch to exponentiated coordinates
\begin{equation}
    z_i:=e^{u_i},\qquad w_i:=-e^{v_i}.
\end{equation}
Making note of the sign in the definition of the $\{w_i\}$, the saddle-point equations then take the form
\begin{subequations}
\begin{align}
    \kappa_{1} \log z_{i} &=\frac{t_{1}}{N_{1}} \sum_{j \neq i}^{N_{1}} \frac{z_i+z_{j}}{z_{i}-z_{j}}-\frac{t_{2}}{N_{2}} \sum_{j=1}^{N_{2}} \frac{z_i+w_{j}}{z_{i}-w_{j}}, \\
    \kappa_{2} \log(- w_{i}) &=\frac{t_{2}}{N_{2}} \sum_{j \neq i}^{N_{2}} \frac{w_i+w_{j}}{w_{i}-w_{j}}-\frac{t_{1}}{N_{1}} \sum_{j=1}^{N_{1}} \frac{w_i+z_{j}}{w_{i}-z_{j}}.
\end{align}
\label{eq:mmspe}
\end{subequations}

We now define the planar resolvent in terms of the exponentiated variables
\begin{equation}
    v(z):=v_1(z)-v_2(z)=\fft{t_1}{N_1}\sum_{i=1}^{N_1}\fft{z+z_i}{z-z_i}-\fft{t_2}{N_2}\sum_{i=1}^{N_2}\fft{z+w_i}{z-w_i},
\label{eq:resolventi}
\end{equation}
where the eigenvalues $\left\{z_i\right\}_{i=1}^{N_1}$ and $\left\{w_i\right\}_{i=1}^{N_2}$ solve the saddle-point equations \eqref{eq:mmspe}. In the planar limit, $k\to\infty$, we expect the eigenvalue distributions $\{z_i\}_{i=1}^{N_1}$ to localize to a cut $[c,d]\subset\mathbb{R}^+$ and 
$\{w_i\}_{i=1}^{N_2}$ to localize to a cut $[a,b]\subset\mathbb{R}^-$.  We thus introduce eigenvalue densities $\rho(x)$ and $\tilde\rho(x)$ and write the planar resolvent as
\begin{equation}
    v(z):=t_1\int_c^d\dd{x}\rho(x)\frac{z+x}{z-x}-t_2\int_a^b\dd{x}\tilde{\rho}(x)\frac{z+x}{z-x}.
\label{eq:expResolvent}
\end{equation}
Note that $v(z)$ has branch-cut discontinuities along $[a,b]$ and $[c,d]$ where the eigenvalues condense.  In terms of this resolvent, we can rewrite the saddle-point equations quite simply as
\begin{subequations}
\begin{align}
    \kappa_{1} \log z& =\ft12[v(z+i 0)+v(z-i 0)],\qquad y\in[c,d] \\
    -\kappa_{2} \log (-z) & =\ft12[v(z+i 0)+v(z-i 0)],\qquad y\in[a,b]
\end{align}\label{eq:expSPE}
\end{subequations}
These equations can be solved by standard methods that have been developed in random matrix theory (see \textit{e.g.}~\cite{Marino:2011nm}).

In fact, the planar resolvent for GT theory was already worked out in \cite{Suyama:2010hr} by solving the Riemann-Hilbert problem.  The idea is to convert the saddle-point equations, (\ref{eq:expSPE}), which correspond to the principal value of the resolvent along the two cuts, into corresponding discontinuity equations by introducing
\begin{equation}
    f(z)=\fft{v(z)}{\sqrt{(z-a)(z-b)(z-c)(z-d)}}.
\end{equation}
We then use Cauchy's theorem to write
\begin{equation}
    f(z)=\oint\fft{d\zeta}{2\pi i}\fft{f(\zeta)}{\zeta-z},
\end{equation}
where the contour is a small circle surrounding $z$.  By deforming the contour to go around the two cuts and using the saddle-point equations, we can obtain an integral expression for $f(z)$.  Converted back to the resolvent, $v(z)$, we finally obtain \cite{Suyama:2010hr}
\begin{align}
    v(z)&=\frac{\kappa_1}{\pi}\int_c^d\dd{x}\frac{\log\qty(x)}{z-x}\frac{\sqrt{(z-a)(z-b)(z-c)(z-d)}}{\sqrt{|(x-a)(x-b)(x-c)(x-d)|}}\nonumber \\
    &\quad+\frac{\kappa_2}{\pi}\int_a^b\dd{x}\frac{\log\qty(-x)}{z-x}\frac{\sqrt{(z-a)(z-b)(z-c)(z-d)}}{\sqrt{|(x-a)(x-b)(x-c)(x-d)|}}.
\label{eq:vzint}
\end{align}
This is the starting point for the subsequent analysis.

\subsection{Fixing the endpoints}

While the GT theory is parametrized by the couplings $t_1$ and $t_2$, the expression (\ref{eq:vzint}) for the resolvent is instead parametrized by the endpoints $a,b,c,d$ of the two cuts.  We thus want to relate these two sets of parameters.  The problem can be simplified by noticing that the saddle-point equations, \eqref{eq:mmspe}, are invariant under $z\to z^{-1}$ and $w\to w^{-1}$. This suggests that the eigenvalue distributions should also be invariant under this map, which leads to an ansatz
\begin{equation}
    ab=1,\ \ cd=1.\label{eq:endpointAnsatz}
\end{equation}
It was shown in \cite{Suyama:2010hr} that this ansatz is consistent with the constraints imposed by the asymptotic behavior of the resolvent $v(z)$ in the limits $z\to\infty$ and $z\to0$.  We still need to relate the two undetermined parameters (say $a$ and $d$) to the couplings $t_1$ and $t_2$.  This can be done using the relations
\begin{subequations}
\begin{align}
    t_1&=\fft1{4\pi i}\oint_{\mathcal{C}_1} {\dd{z}}\frac{v(z)}{z},\\
    t_2&=\fft1{4\pi i}\oint_{\mathcal{C}_2} {\dd{z}}\frac{v(z)}{z},
\end{align}
\label{eq:endpointContour}%
\end{subequations}
which can be derived directly from the expression \eqref{eq:expResolvent} for the resolvent.  Here $\mathcal{C}_1$ and $\mathcal{C}_2$ are contours enclosing the branch cuts $[c,d]$ and $[a,b]$, respectively.

While the resolvent, (\ref{eq:vzint}), does not appear to admit a simple analytic form, we can work with it as an integral expression.  This is facilitated in the strong coupling limit $t_1,t_2\gg1$, where it was shown in \cite{Suyama:2011yz} that the endpoints of the two cuts scale uniformly when $t_1\approx t_2\to\infty$.  In particular, making note of (\ref{eq:endpointAnsatz}), we let
\begin{equation}
    a=-e^\alpha,\qquad b=-e^{-\alpha},\qquad c=e^{-\beta},\qquad d=e^\beta.
\label{eq:abcd}
\end{equation}
Since the strong coupling limit is taken with $\alpha\approx\beta$, we find it convenient to further parametrize the endpoints by
\begin{equation}
    \alpha=\gamma+\delta,\qquad\beta=\gamma-\delta.
\label{eq:abgd}
\end{equation}
The symmetric case, $t_1=t_2$ and $\kappa_1=\kappa_2$, corresponds to $\delta=0$ and
\begin{equation}
    t_1=t_2\sim\frac{\kappa_1+\kappa_2}{3\pi^2}\gamma^3,
\label{eq:endptIntermed}
\end{equation}
at least to leading order \cite{Suyama:2011yz}.  More generally, the scaling $t_i\sim\gamma^3$ continues to hold, while $\delta$ is of subleading order compared with $\gamma$.  The relation between $\{\gamma,\delta\}$ and $\{t_1,t_2\}$ will be worked out in more detail below.

\subsection{Computing the free energy}\label{subsec:freeEnergy}

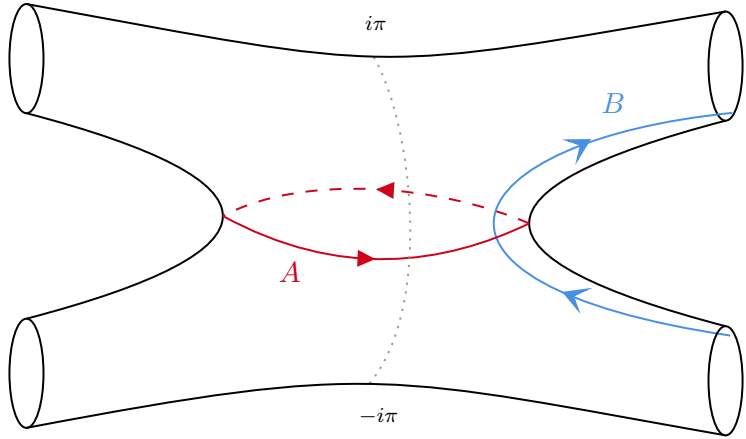
\begin{figure}[t]
    \centering
    \tikzset{every picture/.style={line width=0.75pt}} 
    
    \begin{tikzpicture}[x=0.75pt,y=0.75pt,yscale=-1,xscale=1]
    
    \draw   (455.42,71.6) .. controls (460.12,71.6) and (463.92,83.89) .. (463.92,99.06) .. controls (463.92,114.22) and (460.12,126.52) .. (455.42,126.52) .. controls (450.73,126.52) and (446.92,114.22) .. (446.92,99.06) .. controls (446.92,83.89) and (450.73,71.6) .. (455.42,71.6) -- cycle ;
    \draw   (455.42,126.52) .. controls (324.74,160.97) and (324.74,195.43) .. (455.42,229.89) ;
    \draw   (106.47,226.12) .. controls (237.15,191.66) and (237.15,157.21) .. (106.47,122.75) ;
    \draw [color={rgb, 255:red, 208; green, 2; blue, 27 }  ,draw opacity=1 ] [dash pattern={on 4.5pt off 4.5pt}]  (357.41,178.2) .. controls (313.38,159.27) and (246.42,152.49) .. (204.48,174.43) ;
    \draw [shift={(280.98,161.03)}, rotate = 363.69] [fill={rgb, 255:red, 208; green, 2; blue, 27 }  ,fill opacity=1 ][line width=0.08]  [draw opacity=0] (8.93,-4.29) -- (0,0) -- (8.93,4.29) -- cycle    ;
    \draw [color={rgb, 255:red, 74; green, 144; blue, 226 }  ,draw opacity=1 ]   (458.8,122.6) .. controls (319.3,136.57) and (282.18,208.31) .. (457.8,234.6) ;
    \draw [color={rgb, 255:red, 208; green, 2; blue, 27 }  ,draw opacity=1 ]   (204.48,174.43) .. controls (262.88,206.55) and (314.69,198.93) .. (357.41,178.2) ;
    \draw [shift={(280.64,196.12)}, rotate = 182.38] [fill={rgb, 255:red, 208; green, 2; blue, 27 }  ,fill opacity=1 ][line width=0.08]  [draw opacity=0] (8.93,-4.29) -- (0,0) -- (8.93,4.29) -- cycle    ;
    \draw  [color={rgb, 255:red, 74; green, 144; blue, 226 }  ,draw opacity=1 ][fill={rgb, 255:red, 74; green, 144; blue, 226 }  ,fill opacity=1 ] (381.64,221.32) -- (374.73,213) -- (385.96,211.5) -- (379.26,214.7) -- cycle ;
    \draw  [color={rgb, 255:red, 74; green, 144; blue, 226 }  ,draw opacity=1 ][fill={rgb, 255:red, 74; green, 144; blue, 226 }  ,fill opacity=1 ] (376.62,136.42) -- (387.43,136.24) -- (381.57,145.93) -- (383.26,138.7) -- cycle ;
    \draw   (455.42,229.89) .. controls (460.12,229.89) and (463.92,242.18) .. (463.92,257.34) .. controls (463.92,272.51) and (460.12,284.8) .. (455.42,284.8) .. controls (450.73,284.8) and (446.92,272.51) .. (446.92,257.34) .. controls (446.92,242.18) and (450.73,229.89) .. (455.42,229.89) -- cycle ;
    \draw   (106.47,226.12) .. controls (111.16,226.12) and (114.97,238.41) .. (114.97,253.57) .. controls (114.97,268.74) and (111.16,281.03) .. (106.47,281.03) .. controls (101.77,281.03) and (97.97,268.74) .. (97.97,253.57) .. controls (97.97,238.41) and (101.77,226.12) .. (106.47,226.12) -- cycle ;
    \draw   (106.47,67.83) .. controls (111.16,67.83) and (114.97,80.12) .. (114.97,95.29) .. controls (114.97,110.45) and (111.16,122.75) .. (106.47,122.75) .. controls (101.77,122.75) and (97.97,110.45) .. (97.97,95.29) .. controls (97.97,80.12) and (101.77,67.83) .. (106.47,67.83) -- cycle ;
    \draw [color={rgb, 255:red, 155; green, 155; blue, 155 }  ,draw opacity=1 ] [dash pattern={on 0.84pt off 2.51pt}]  (279.8,94.6) .. controls (298.8,126.6) and (309.8,219.6) .. (277.8,258.6) ;
    \draw    (106.47,67.83) .. controls (294.8,102.6) and (275.8,102.6) .. (455.42,71.6) ;
    \draw    (106.47,281.03) .. controls (281.8,247.6) and (285.8,254.6) .. (455.42,284.8) ;
    
    \draw (392,111.4) node [anchor=north west][inner sep=0.75pt]    {$\textcolor[rgb]{0.29,0.56,0.89}{B}$};
    \draw (231,196.4) node [anchor=north west][inner sep=0.75pt]    {$\textcolor[rgb]{0.82,0.01,0.11}{A}$};
    \draw (271,269.4) node [anchor=north west][inner sep=0.75pt]  [font=\scriptsize]  {$-i\pi $};
    \draw (274,72.4) node [anchor=north west][inner sep=0.75pt]  [font=\scriptsize]  {$i\pi $};

    \end{tikzpicture}

    \caption{The $A$ and $B$-cycle contours for Chern-Simons theory. Note that the Riemann sheets are curled up due to the $2\pi i$ periodicity of the resolvent.}
    \label{fig:Bcycle}
\end{figure}

While the leading order free energy, (\ref{eq:FGT0}), can be obtained directly from a large-$N$ saddle point solution \cite{Jafferis:2011zi}, since we are interested in subleading corrections, we will instead work with the resolvent, following \cite{Suyama:2010hr,Suyama:2011yz}.  In particular, making the identification $g_s={2\pi i}/{k}$, the free energy can be written in the form of a genus expansion
\begin{equation}
    F=\sum_{g=0}^\infty g_s^{2g-2}F_g(t).
\end{equation}
It has long been known that the genus-zero free energy, $F_0(t)$, for such matrix models can be written as an integral of the planar resolvent over a particular contour \cite{Dijkgraaf:2002fc,Halmagyi:2003fy,Halmagyi:2003ze}%
\footnote{Note that the convention for the free energy in this paper differs from that in \cite{Halmagyi:2003fy,Halmagyi:2003ze} in that we take the free energy to be $F=-\log Z$.}.
The basic idea is to look at the change in the leading order free energy from adding one eigenvalue to the branch cut, and use this to deduce the derivative of the genus-zero free energy with respect to the 't~Hooft parameter. The resulting expression can then be shown to an integral of the resolvent around the $B$-cycle, a contour that starts at infinity on one Riemann sheet, passes through the branch cut, and goes off to infinity on the other Riemann sheet. This results in a beautiful geometric picture, where the $A$-cycle determines the endpoints and the $B$-cycle determines the free energy; this is depicted in Figure \ref{fig:Bcycle} for the Chern-Simons matrix model.  This is the strategy we will employ for GT theory.

\begin{figure}[t]
    \centering
    \tikzset{every picture/.style={line width=0.75pt}} 
    
    \begin{tikzpicture}[x=0.75pt,y=0.75pt,yscale=-1,xscale=1]
    
    \draw   (274.46,156.34) .. controls (274.46,128.56) and (298.36,106.04) .. (327.85,106.04) .. controls (357.33,106.04) and (381.23,128.56) .. (381.23,156.34) .. controls (381.23,184.12) and (357.33,206.64) .. (327.85,206.64) .. controls (298.36,206.64) and (274.46,184.12) .. (274.46,156.34) -- cycle ;
    \draw [color={rgb, 255:red, 208; green, 2; blue, 27 }  ,draw opacity=1 ]   (178.28,152.81) .. controls (195.55,163.46) and (247.35,169.2) .. (274.46,152.45) ;
    \draw [shift={(226.58,162.99)}, rotate = 181.65] [fill={rgb, 255:red, 208; green, 2; blue, 27 }  ,fill opacity=1 ][line width=0.08]  [draw opacity=0] (8.93,-4.29) -- (0,0) -- (8.93,4.29) -- cycle    ;
    \draw [color={rgb, 255:red, 208; green, 2; blue, 27 }  ,draw opacity=1 ]   (381.23,154.74) .. controls (398.49,165.39) and (450.29,171.13) .. (477.41,154.38) ;
    \draw [shift={(429.52,164.93)}, rotate = 181.65] [fill={rgb, 255:red, 208; green, 2; blue, 27 }  ,fill opacity=1 ][line width=0.08]  [draw opacity=0] (8.93,-4.29) -- (0,0) -- (8.93,4.29) -- cycle    ;
    \draw  [color={rgb, 255:red, 74; green, 144; blue, 226 }  ,draw opacity=1 ][fill={rgb, 255:red, 74; green, 144; blue, 226 }  ,fill opacity=1 ] (152.56,223.76) -- (163.79,222.39) -- (159.27,232.17) -- (159.85,225.18) -- cycle ;
    \draw  [color={rgb, 255:red, 74; green, 144; blue, 226 }  ,draw opacity=1 ][fill={rgb, 255:red, 74; green, 144; blue, 226 }  ,fill opacity=1 ] (171.55,98.21) -- (168.56,87.92) -- (179.47,90.79) -- (172.04,91.21) -- cycle ;
    \draw  [color={rgb, 255:red, 74; green, 144; blue, 226 }  ,draw opacity=1 ][fill={rgb, 255:red, 74; green, 144; blue, 226 }  ,fill opacity=1 ] (499.41,72.01) -- (510.67,73.06) -- (503.93,81.63) -- (506.17,74.94) -- cycle ;
    \draw  [color={rgb, 255:red, 74; green, 144; blue, 226 }  ,draw opacity=1 ][fill={rgb, 255:red, 74; green, 144; blue, 226 }  ,fill opacity=1 ] (513.35,242.65) -- (508.13,233.18) -- (519.44,233.82) -- (512.26,235.7) -- cycle ;
    \draw    (27.5,45.04) .. controls (220.94,44.99) and (238.2,260.3) .. (26.48,260.58) ;
    \draw [color={rgb, 255:red, 74; green, 144; blue, 226 }  ,draw opacity=1 ]   (27.5,45.04) .. controls (258.52,45.95) and (263.59,262.21) .. (26.48,260.58) ;
    \draw    (629.21,47.57) .. controls (422.03,49.77) and (430.15,258.38) .. (629.21,262.15) ;
    \draw [color={rgb, 255:red, 74; green, 144; blue, 226 }  ,draw opacity=1 ]   (629.21,47.61) .. controls (394.61,46.9) and (397.65,261.25) .. (629.21,262.19) ;
    \draw [color={rgb, 255:red, 208; green, 2; blue, 27 }  ,draw opacity=1 ] [dash pattern={on 4.5pt off 4.5pt}]  (477.41,154.38) .. controls (441.71,139.48) and (415.23,137.47) .. (381.23,154.74) ;
    \draw [shift={(429.14,142.51)}, rotate = 363] [fill={rgb, 255:red, 208; green, 2; blue, 27 }  ,fill opacity=1 ][line width=0.08]  [draw opacity=0] (8.93,-4.29) -- (0,0) -- (8.93,4.29) -- cycle    ;
    \draw [color={rgb, 255:red, 208; green, 2; blue, 27 }  ,draw opacity=1 ] [dash pattern={on 4.5pt off 4.5pt}]  (274.46,152.45) .. controls (238.76,137.55) and (212.28,135.54) .. (178.28,152.81) ;
    \draw [shift={(226.2,140.57)}, rotate = 363] [fill={rgb, 255:red, 208; green, 2; blue, 27 }  ,fill opacity=1 ][line width=0.08]  [draw opacity=0] (8.93,-4.29) -- (0,0) -- (8.93,4.29) -- cycle    ;
    
    \draw (170.97,61.23) node [anchor=north west][inner sep=0.75pt]  [color={rgb, 255:red, 74; green, 144; blue, 226 }  ,opacity=1 ]  {$B_{2}$};
    \draw (454.92,73.89) node [anchor=north west][inner sep=0.75pt]  [color={rgb, 255:red, 74; green, 144; blue, 226 }  ,opacity=1 ]  {$B_{1}$};
    \draw (416.36,118.08) node [anchor=north west][inner sep=0.75pt]  [color={rgb, 255:red, 208; green, 2; blue, 27 }  ,opacity=1 ]  {$\mathcal{C}_{1}$};
    \draw (224.91,117.38) node [anchor=north west][inner sep=0.75pt]  [color={rgb, 255:red, 208; green, 2; blue, 27 }  ,opacity=1 ]  {$\mathcal{C}_{2}$};
    \end{tikzpicture}
    \caption{The integration contours (in exponentiated coordinates) used in the derivation of the genus-zero free energy. Note that we no longer have the $2\pi i$ periodicity of the Riemann sheets because we are in exponentiated coordinates.}
    \label{fig:contours}
\end{figure}
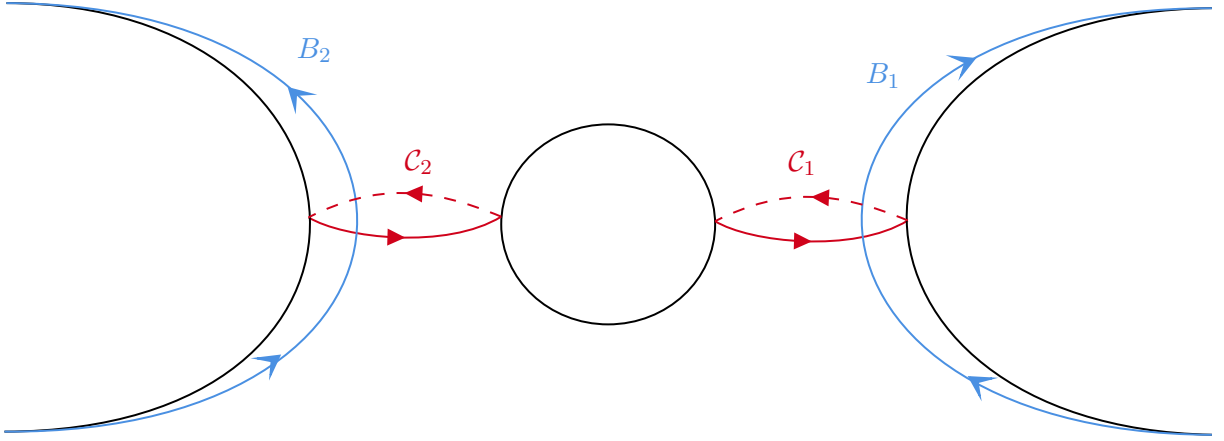

The two-node GT theory has two gauge groups whose eigenvalues condense along separate cuts in the complex plane.  As a result, there are two B-cycle integrals to consider.  We start by taking the genus-zero free energy $F_0=\left.g_s^2S\right|_{N\to\infty}$ from the effective action, (\ref{eq:action}).  For the first gauge group, we play the trick of adding one more $\hat u$ eigenvalue to the first branch cut (\textit{i.e.} we take $N_1\to N_1+1$).  The 't~Hooft parameter correspondingly changes by $\Delta t_1={2\pi}/{k}$. This gives
\begin{equation}
    \fft{\Delta F_0}{\Delta t_1}=\fft{\kappa_1}2\hat u^2-t_1\fft1{N_1}\sum_i^{N_1}\log\sinh^2\fft{\hat u-u_i}2+t_2\fft1{N_2}\sum_i^{N_2}\log\cosh^2\fft{\hat u-v_i}2.
\end{equation}
Integrating the resolvent, (\ref{eq:resolventi})
\begin{equation}
    v_1(z)=\fft{t_1}{N_1}\sum_{i=1}^{N_1}\fft{z+z_i}{z-z_i},
\end{equation}
we then obtain
\begin{equation}
    \fft{t_1}{N_1}\sum_{i=1}^{N_1}\log\sinh^2\fft{\hat u-u_i}2=-\int_{e^{\hat u}}^{e^\Lambda}v_1(z)\fft{dz}z+t_1(\Lambda-\log4),
\end{equation}
where $\Lambda$ is a large cutoff and we have dropped exponentially small terms of the form $e^{-\Lambda}$.  Using this expression and a similar one for the integral of $v_2(z)$ gives, in the large-$N$ limit
\begin{equation}
    \fft{\partial F_0}{\partial t_1}=\fft{\kappa_1}2\hat u^2+\int_{e^{\hat u}}^{e^\Lambda}v(z)\fft{dz}z-(t_1-t_2)(\Lambda-\log4).
\end{equation}
We take the last eigenvalue $\hat u$ at the right endpoint of the cut, (\ref{eq:abcd}), and write
\begin{equation}
    \fft{\partial F_0}{\partial t_1}=\fft{\kappa_1}2\beta^2+\int_\beta^\Lambda v(e^u)du-(t_1-t_2)(\Lambda-\log4).
\label{eq:dF0dt1}
\end{equation}
Geometrically, this is the $B_1$-cycle integral, which we have graphically depicted in Figure~\ref{fig:contours}.  By swapping the two gauge groups, we can obtain a similar $B_2$-cycle integral for $\partial F_0/\partial t_2$.  This integral will be worked out perturbatively in the next section.

\section{Subleading Corrections to the Free Energy}
\label{sec:GTfree}

We now turn to an evaluation of the free energy beyond leading order.  As we have seen above in (\ref{eq:vzint}), the planar resolvent for the GT model can be written down in integral form.  While the integral is challenging to perform analytically, the general expression will be sufficient when working out the free energy.

Our goal is to compute the derivative of the free energy, (\ref{eq:dF0dt1}), up to exponentially small terms in the large 't~Hooft parameter limit.  To do so, we insert the integral expression for the resolvent, (\ref{eq:vzint}), into (\ref{eq:dF0dt1}) and work out the double integral in the large $t_1$ and $t_2$ limit.  However, since this gives an expression for $\partial F_0/\partial t_1$ as a function of the endpoints of the cuts, (\ref{eq:abcd}), we additionally need to relate the endpoints to the 't~Hooft couplings using the $A$-cycle integrals (\ref{eq:endpointContour}).  We will work this out first and then return to the free energy integral.

\subsection{Correction to the endpoints}

At leading order, the endpoints of the cuts scale with the 't~Hooft couplings according to (\ref{eq:endptIntermed}).  However this will pick up corrections, both for $t_1\ne t_2$ and subleading in the couplings.  We explicitly work out the $A$-cycle integral for $t_1$; then the $t_2$ expression follows from symmetry under $t_1\leftrightarrow t_2$ and $\kappa_1\leftrightarrow\kappa_2$ interchange.

Substituting the integral expression for the resolvent, (\ref{eq:vzint}), into (\ref{eq:endptIntermed}), then explicitly writing out the $A$-cycle integral as an integral over the discontinuity across the cut and finally interchanging the order of integration gives
\begin{equation}
    t_1=\fft{\kappa_1}{2\pi^2}J_1+\fft{\kappa_2}{2\pi^2}J_2,
\end{equation}
where
\begin{subequations}
\begin{align}
    J_1=\int_c^ddx\fft{\log x}{\sqrt{(x-a)(x-b)(x-c)(d-x)}}I(x),\\ J_2=\int_a^bdx\fft{\log(-x)}{\sqrt{(x-a)(b-x)(c-x)(d-x)}}I(x),
\end{align}
\label{eq:J1J2ints}%
\end{subequations}
with
\begin{equation}
    I(z)=\int_c^d\fft{dy}y\fft{\sqrt{(y-a)(y-b)(y-c)(d-y)}}{z-y}.
\end{equation}
Here the principal value of $I(x)$ has to be taken in the $J_1$ integral.  We proceed by rewriting these expressions in terms of exponentiated variables:
\begin{subequations}
\begin{align}
    J_1&=\int_{-\beta}^\beta dv\fft{vI(e^v)}{4\sqrt{\cosh\fft{\alpha+v}2\cosh\fft{\alpha-v}2\sinh\fft{\beta+v}2\sinh\fft{\beta-v}2}},\\
    J_2&=\int_{-\alpha}^\alpha dv\fft{vI(-e^v)}{4\sqrt{\sinh\fft{\alpha+v}2\sinh\fft{\alpha-v}2\cosh\fft{\beta+v}2\cosh\fft{\beta-v}2}},
\end{align}
\end{subequations}
and
\begin{equation}
    I(z)=\int_{-\beta}^\beta du\fft{4\sqrt{\cosh\fft{\alpha+u}2\cosh\fft{\alpha-u}2\sinh\fft{\beta+u}2\sinh\fft{\beta-u}2}}{ze^{-u}-1}.
\end{equation}
Note that the $\cosh$ terms are never vanishing, while the $\sinh$ terms vanish at the endpoints.  Moreover, the square-root factors are all even under $v\to-v$ or $u\to-u$.  This suggests that we split up the regions of integration into half intervals and write
\begin{subequations}
\begin{align}
    J_1&=\int_0^\beta dv\fft{vI_1(v)}{4\sqrt{\cosh\fft{\alpha+v}2\cosh\fft{\alpha-v}2\sinh\fft{\beta+v}2\sinh\fft{\beta-v}2}},\\
    J_2&=\int_0^\alpha dv\fft{vI_2(v)}{4\sqrt{\sinh\fft{\alpha+v}2\sinh\fft{\alpha-v}2\cosh\fft{\beta+v}2\cosh\fft{\beta-v}2}},
\end{align}
\end{subequations}
where
\begin{subequations}
\begin{align}
    I_1(v)&=\int_0^\beta du\,4\sqrt{\textstyle\cosh\fft{\alpha+u}2\cosh\fft{\alpha-u}2\sinh\fft{\beta+u}2\sinh\fft{\beta-u}2}\left(\coth\fft{v-u}2+\coth\fft{v+u}2\right),\\
    I_2(v)&=\int_0^\beta du\,4\sqrt{\textstyle\cosh\fft{\alpha+u}2\cosh\fft{\alpha-u}2\sinh\fft{\beta+u}2\sinh\fft{\beta-u}2}\left(\tanh\fft{v-u}2+\tanh\fft{v+u}2\right).
\end{align}
\end{subequations}
Here we see explicitly that the integrand of $I_1$ has a pole when $v-u$ vanishes, so the principal value ought to be taken when evaluating the integral.

So far, these expressions are still exact, as far as the planar resolvent is concerned.  However, the integrals are not easy to evaluate.  To proceed, we now focus on the large 't~Hooft coupling limit, where $\alpha,\beta\gg1$.  Since the integrals are over half intervals, we can approximate $\alpha+v\gg1$, $\beta+v\gg1$ and similarly for $v$ replaced by $u$.  As a result, up to exponentially suppressed terms, we have
\begin{align}
    J_1=\int_0^\beta dv\fft{ve^{-\fft12(\gamma+v)}I_1(v)}{2\sqrt{\cosh\fft{\alpha-v}2\sinh\fft{\beta-v}2}},\qquad
    J_2=\int_0^\alpha dv\fft{ve^{-\fft12(\gamma+v)}I_2(v)}{2\sqrt{\sinh\fft{\alpha-v}2\cosh\fft{\beta-v}2}},
\label{eq:J1intdef}
\end{align}
with
\begin{subequations}
\begin{align}
    I_1(v)&=\int_0^\beta du\,2e^{\fft12(\gamma+u)}\sqrt{\textstyle\cosh\fft{\alpha-u}2\sinh\fft{\beta-u}2}\left(\coth\fft{v-u}2+\coth\fft{v+u}2\right),\\
    I_2(v)&=\int_0^\beta du\,2e^{\fft12(\gamma+u)}\sqrt{\textstyle\cosh\fft{\alpha-u}2\sinh\fft{\beta-u}2}\left(\tanh\fft{v-u}2+\tanh\fft{v+u}2\right).
\end{align}
\label{eq:I12intdef}%
\end{subequations}
Recall that we have defined $\gamma=(\alpha+\beta)/2$ and $\delta=(\alpha-\beta)/2$, following (\ref{eq:abcd}).

The $I_1$ and $I_2$ integrals can be performed explicitly and then substituted into the integrands for $J_1$ and $J_2$.  The remaining integrals are more challenging, and we have been unable to obtain a closed form expression for $J_1$ and $J_2$.  Nevertheless, they can be reduced to polynomial expressions in $\gamma$ up to exponentially suppressed terms.  The integration is worked out in Appendix~\ref{appendix:endpoints}, and the result is a relation between the 't~Hooft couplings $t_1$ and $t_2$ and the endpoints of the cuts as parametrized by $\gamma$ and $\delta$.  After defining convenient combinations of $t_1$ and $t_2$,
\begin{equation}
    \bar t=\ft12(t_1+t_2),\qquad\Delta=\ft12(t_1-t_2),
\end{equation}
we find
\begin{subequations}
\begin{align}
    \bar t&=\fft{\kappa_1+\kappa_2}{4\pi^2}\biggl[\fft43(\gamma-\log\ft12\cosh\delta)^3+4\gamma\tan^{-1}\sinh\delta(\tan^{-1}\sinh\delta-\xi)\nn\\
    &\kern5em+\fft43\log^3(\ft12\cosh\delta)+j_{1,e}(\delta)+j_{2,e}(\delta)+\fft{2\xi}\pi j_{1,o}(\delta)\biggr],\\
    \Delta&=\fft{\kappa_1+\kappa_2}{4\pi^2}\Bigl[-2\pi\gamma(\tan^{-1}\sinh\delta-\xi)+j_{1,o}(\delta)+2\pi \xi\log\ft12\cosh\delta\Bigr],
\end{align}
\label{eq:tbD}%
\end{subequations}
where
\begin{equation}
    \xi:=\fft\pi2\fft{\kappa_1-\kappa_2}{\kappa_1+\kappa_2}=\fft\pi2\fft{k_1-k_2}{k_1+k_2},
\label{eq:xdef}
\end{equation}
is the relative difference in Chern-Simons levels.  Here $j_1(\delta)$ and $j_2(\delta)$ are particular functions explicitly defined in Appendix \ref{appendix:endpoints}, and the subscripts $e$ and $o$ denote their even and odd parts, respectively.

For the most part, we are interested in the case of equal ranks, $N_1=N_2$, in which case the difference $\Delta$ vanishes.  Setting $\Delta=0$ in (\ref{eq:tbD}) then gives a straightforward expression for $\gamma$ in terms of $\delta$
\begin{equation}
    2\pi\gamma(\tan^{-1}\sinh\delta-\xi)=j_{1,o}(\delta)+2\pi \xi\log\ft12\cosh\delta.
\label{eq:2pgeqn}
\end{equation}
However, we are actually more interested in obtaining $\delta$ in terms of $\gamma$ since we are focused on the large coupling expansion generalizing (\ref{eq:endptIntermed}) where $t_i\sim\gamma^3$ with $\delta$ being subdominant.  Working to leading order in $\gamma$, we can disregard the last two terms in the expression for $\Delta$ in (\ref{eq:tbD}), so that
\begin{equation}
    \delta\approx\sinh^{-1}\tan\left(\xi\right).
\end{equation}
However, we can do better than this. Since we assume $\gamma\gg1$, we can expand perturbatively
\begin{equation}
    \delta\approx\sinh^{-1}\tan\left(\xi\right)+\frac{\delta_1}{\gamma}+\frac{\delta_2}{\gamma^2}+\cO\qty(\frac{1}{\gamma^3}).
\label{eq:delta}
\end{equation}
Solving the $\cO(\gamma^0)$ expression in $\Delta$ gives
\begin{equation}
    \delta_1=\sec \left(\xi\right)\Cl_2\left(\pi+2\xi\right),
\end{equation}
where $\Cl_2(x)$ denotes the Clausen function
\begin{equation}
    \Cl_2(x)=\Im\Li_2(e^{ix}).
\end{equation}
The expression for $\delta_2$ is rather more involved
\makeatletter
\newcommand{\vast}{\bBigg@{4}}
\makeatother
\begin{align}
    \delta_2&=\frac{1}{2}\sec\qty(\xi)\Cl_2\qty(\pi+2\xi)
    \vast[\tan \left(\xi\right) \left(2\xi-2 \text{gd}\left(\sinh ^{-1}\left(\tan \left(\xi\right)\right)\right)+\Cl_2\qty(\pi+2\xi)\right)\nn\\
    &-4 \sinh ^{-1}\left(\tan \left(\xi\right)\right)-2 \log \left(\frac{8\sec\qty(\xi)}{\left(\qty(\tan \left(\xi\right)+\sec \left(\xi\right))^2+1\right){}^2}\right)\vast],
\end{align}
where $\text{gd}$ denotes the Gudermannian function
\begin{equation}
    \mathrm{gd}(z)=2\arctan\tanh\qty(\tfrac{1}{2}z).
\end{equation}
This is a rather messy expression, but for $\xi\ll 1$, it takes the nice perturbative form
\begin{equation}
    \delta_2\approx 2\xi \log ^2(2)+\xi^3\left(3\log ^2(2)-\fft43 \log (2)\right)+\mathcal{O}\qty(\xi^5).
\end{equation}

Having obtained $\delta$, at least perturbatively, we now proceed to related $\gamma$ and $\bar t$.  Keeping $\Delta=0$, we first eliminate the second term in the $\bar t$ expression in (\ref{eq:tbD}) to obtain
\begin{align}
    \bar t&=\fft{\kappa_1+\kappa_2}{4\pi^2}\biggl[\fft43(\gamma-\log\ft12\cosh\delta)^3+\fft2\pi(\tan^{-1}\sinh\delta+\xi)j_{1,o}(\delta)\nn\\
    &\kern5em+4\log(\ft12\cosh\delta)\left(\fft13\log^2(\ft12\cosh\delta)+\xi\tan^{-1}\sinh\delta\right)
    +j_{1,e}(\delta)+j_{2,e}(\delta)\biggr].
\end{align}
This expression is useful since now the only $\gamma$ dependence shows up in the first term.  We can now substitute the perturbative expression (\ref{eq:delta}).  To first non-trivial order, we find
\begin{align}
    \bar t&=\fft{\kappa_1+\kappa_2}{4\pi^2}\biggl[\fft43(\gamma-\log\ft12\cosh\delta)^3-4\log(2\cos \xi)\left(\fft13\log^2(2\cos \xi)+\xi^2\right)\nn\\
    &\kern5em+\fft{4\xi}\pi j_{1,o}(\delta)+j_{1,e}(\delta)+j_{2,e}(\delta)+\mathcal O(\gamma^{-1})\biggr].
\end{align}
The transcendental functions on the second line are a bit troublesome to work with.  However, by studying the series expansion of $j_1(\delta)$ and $j_2(\delta)$, we can determine empirically that
\begin{equation}
    \bar t=\fft{\kappa_1+\kappa_2}{4\pi^2}\left[\fft43(\gamma-\log\ft12\cosh\delta)^3-4(\Cl_3(\pi-2\xi)+\zeta(3))+\mathcal O(\gamma^{-1})\right],
\label{eq:tbgeqn}
\end{equation}
where
\begin{equation}
    \Cl_3(x)=\Re\Li_3(e^{ix}).
\end{equation}
We will make use of this expression below when computing the planar free energy.

\subsection{The free energy}

We now turn to the evaluation of the free energy which can be obtained from the integral expression, (\ref{eq:dF0dt1}).  The $B$-cycle integral can be evaluated in a similar manner as the $A$-cycle integral performed above for computing the endpoint relation.  In particular using the integral expression for the resolvent, (\ref{eq:vzint}), we can write
\begin{equation}
    \fft{\partial F_0}{\partial t_1}=\fft{\kappa_1}2\beta^2-(t_1-t_2)(\Lambda-\log4)-\fft{\kappa_1}\pi K_1-\fft{\kappa_2}\pi K_2,
\end{equation}
where
\begin{subequations}
\begin{align}
    K_1=\int_c^ddx\fft{\log x}{\sqrt{(x-a)(x-b)(x-c)(d-x)}}I_B(x),\\ K_2=\int_a^bdx\fft{\log(-x)}{\sqrt{(x-a)(b-x)(c-x)(d-x)}}I_B(x).
\end{align}
\end{subequations}
These integrals are similar to the $J_1$ and $J_2$ integrals in (\ref{eq:J1J2ints}), except that now $I_B(x)$ is a $B$-cycle integral
\begin{equation}
    I_B(z)=\int_d^{e^\Lambda}\fft{dy}y\fft{\sqrt{(y-a)(y-b)(y-c)(y-d)}}{z-y}.
\end{equation}
These integrals can be evaluated up to exponentially small terms in a similar manner as was done for the endpoint integrals.  Combining $\partial F_0/\partial t_1$ and the corresponding expression for $\partial F_0/\partial t_2$, we find the relatively compact expression
\begin{equation}
    \fft{\partial F_0}{\partial\bar t}=\fft{\kappa_1+\kappa_2}2\Bigl[(\gamma-\log\ft12\cosh\delta)^2+(\tan^{-1}\sinh\delta-\xi)^2-\ft1{12}\pi^2-\xi^2\Bigr].
\label{eq:dF0dtb}
\end{equation}
Details of the calculation are given in Appendix \ref{appendix:freeEnergy}.

We now have everything we need to obtain the planar free energy from the resolvent.  Since the derivative $\partial F_0/\partial\bar t$ is given in terms of the endpoint parameters $\gamma$ and $\delta$, the general procedure is to first obtain these parameters from the 't~Hooft couplings $t_1$ and $t_2$ by inverting the endpoint relations (\ref{eq:tbD}).  After doing so, it becomes straightforward to integrate (\ref{eq:dF0dtb}) to obtain the planar free energy $F_0$ up to a $\bar t$ independent constant which remains to be fixed.

Focusing on the case $\Delta=0$, the relation (\ref{eq:2pgeqn}) demonstrates that the combination $(\tan^{-1}\sinh\delta-\xi)$ is of $\mathcal O(\gamma^{-1})$.  As a result, (\ref{eq:dF0dtb}) can be written as
\begin{equation}
     \fft{\partial F_0}{\partial\bar t}=\fft{\kappa_1+\kappa_2}2\Bigl[(\gamma-\log\ft12\cosh\delta)^2-\ft1{12}\pi^2-\xi^2+\mathcal O(\gamma^{-2})\Bigr].
\end{equation}
Making use of the $\bar t$ versus $\gamma$ relation, (\ref{eq:tbgeqn}), and integrating then gives the genus zero free energy
\begin{equation}
    F_0=\bar\kappa^2\left[\fft35\left(\fft{3\pi^2}2\right)^{2/3}\left(\fft{\bar t}{\bar\kappa}+\fft2{\pi^2}\left(\Cl_3(\pi-2\xi)+\zeta(3)\right)\right)^{5/3}-\left(\fft{\pi^2}{12}+\xi^2\right)\fft{\bar t}{\bar\kappa}+\mathcal O(\bar t^{1/3})+\mbox{const.}\right],
\label{eq:f0fin}
\end{equation}
where we have defined
\begin{equation}
    \bar\kappa=\fft{\kappa_1+\kappa_2}2=\fft{k_1+k_2}{2k}.
\end{equation}
Several points are now in order.  Firstly, the ``constant'' term is independent of $\bar t$ but can depend on the fractional difference of Chern-Simons levels, $\xi$.  However, it cannot be obtained directly from integrating the derivative of the free energy%
\footnote{In contrast, the $\cO(\ol{t}^{1/3})$ part can, in principle, be obtained term-by-term from higher-order perturbation theory. We denote the $\cO(\ol{t}^{1/3})$ and constant terms separately to emphasize this distinction.}.
In addition, the leading term in the large-$\bar t$ expansion of this expression matches what we expect from \eqref{eq:FGT=}.  Finally, note that the $\mathcal O(\bar t^{1/3})$ term vanishes in the $\xi=0$ limit (ie, for $k_1=k_2$).  In this case, expression (\ref{eq:f0fin}) is exact up to exponentially small terms in $\bar t$.

In fact, it is easily seen that $\delta$ vanishes in the $\xi=0$ limit.  As a result, (\ref{eq:tbgeqn}) takes on the simple relation
\begin{equation}
    t=\fft{\kappa}{2\pi^2}\left(\ft43(\gamma+\log2)^3-\zeta(3)\right),
\end{equation}
and the planar free energy, (\ref{eq:f0fin}) becomes
\begin{equation}
    F_0=\kappa^2\left[\fft35\left(\fft{3\pi^2}2\right)^{2/3}\left(\frac{t}{\kappa}+\fft{\zeta(3)}{2\pi^2}\right)^{5/3}-\fft{\pi^2}{12}\frac{t}{\kappa}+\mbox{const.}+\mathcal O(e^{-t})\right].
\label{eq:F0equal}
\end{equation}
Here we have dropped the bars on $t$ and $\kappa$ as we are considering $t_1=t_2$ and $\kappa_1=\kappa_2$.  If desired, this can be expanded in inverse powers of $t$
\begin{equation}
    F_0(t)=-\frac{\pi^2}{12} t+\frac{3\cdot6^{2/3}}{40\pi^2}\qty(2\pi^2{t})^{5/3}\sum_{n=0}^\infty \frac{\qty(\tfrac{5}{3})_n}{n!}\qty(\frac{\zeta(3)}{2\pi^2t})^n,
\label{eq:F0expand}
\end{equation}
where $(\ )_n$ denotes the Pochhammer symbol.  Since this expression holds for $k_1=k_2$, we have set $\kappa=1$ and $t=2\pi iN/k$.  Note that $t$ is imaginary when we take $N$ and $k$ to be real.  In this case, the first term, which is linear in $t$, does not contribute to the real part of the free energy.

\subsection{Numerical Analysis}

Our main result is the expression, (\ref{eq:f0fin}), for the genus zero free energy $F_0(N,k_1,k_2)$ at large 't~Hooft coupling $\bar t$.  While the first term is complete, additional terms of $\mathcal O(\bar t^{1/3})$ and smaller will contribute when the Chern-Simons levels are different, as parametrized by $\xi$ defined in (\ref{eq:xdef}).  In order to get an idea of the size of these terms, we carried out a numerical investigation of the large-$N$ partition function.  In this limit, we solved the saddle-point equations in Mathematica for $N$ ranging from 100 to 340 at fixed (real positive) 't~Hooft coupling $\bar t$ and extrapolated $N\to \infty$, using a working precision of 50. This was done for various values of $\bar t$ and then fitted to extract the subleading coefficients $f_1(\xi)$ and $f_2(\xi)$ in the expansion
\begin{equation}
    F_0(\bar t,\bar\kappa,\xi)=\fft35\left(\fft{3\pi^2}2\right)^{2/3}\bar t^{5/3}\bar\kappa^{1/3}+f_1(\xi)\bar t+f_2(\xi)\bar t^{2/3}+\cdots,
\label{eq:F0exp...}
\end{equation}
Throughout these fits, we hold $\bar\kappa=1$ fixed since changing the value of $\bar\kappa$ is equivalent to an overall rescaling of $k$.  The coefficients $f_1(\xi)$ and $f_2(\xi)$ are then extracted from the numerical free energy for various values of $\xi=(\pi/4)(\kappa_1-\kappa_2)$. Due to the computational difficulty of this process, this was only done for five sample points corresponding to $\xi=\{0,\tfrac{\pi}{40},\tfrac{\pi}{20},\tfrac{\pi}{10},\tfrac{3\pi}{20}\}$.

\begin{figure}[t]
\includegraphics[width=0.75\textwidth]{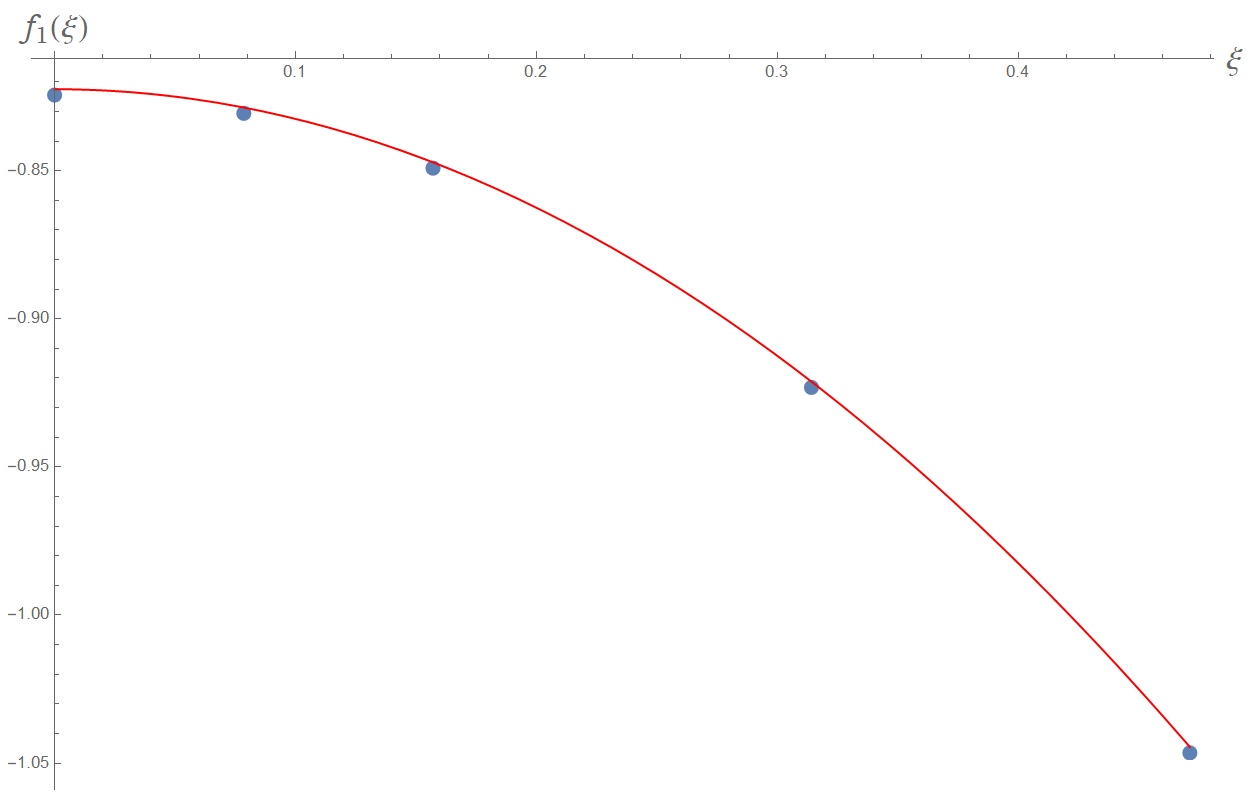}
\centering
\caption{Plot of the coefficient $f_1(\xi)$. The red line is the analytic prediction from \eqref{eq:f0fin}, and the blue dots are sample points for numerical simulations performed in Mathematica for $\xi=0$, $\tfrac{\pi}{40}$, $\tfrac{\pi}{20}$, $\tfrac{\pi}{10}$, and $\tfrac{3\pi}{20}$.}
\label{fig:linearTerm}
\end{figure}

\begin{figure}[t]
\includegraphics[width=0.75\textwidth]{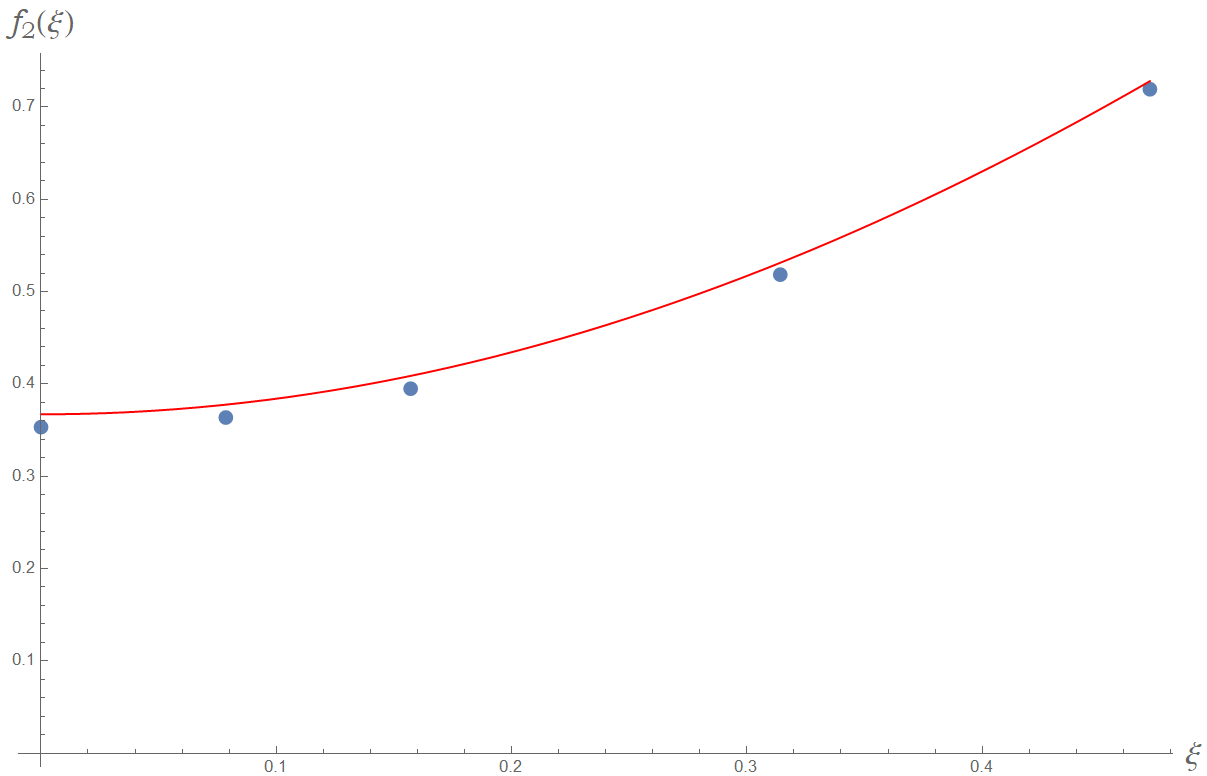}
\centering
\caption{Plot of the coefficient $f_2(\xi)$. The red line is the analytic prediction from \eqref{eq:f0fin}, and the blue dots are sample points for numerical simulations performed in Mathematica for $\xi=0$, $\tfrac{\pi}{40}$, $\tfrac{\pi}{20}$, $\tfrac{\pi}{10}$, and $\tfrac{3\pi}{20}$.}
\label{fig:t2/3term}
\end{figure}

We have verified that the leading order term in \eqref{eq:F0exp...} is reproduced numerically to very high precision and that no term of $\mathcal O(\bar t^{4/3})$ shows up within numerical uncertainties.  As a result, we subtracted the analytic value of the leading term and fit only the subdominant coefficients.  The coefficient $f_1(\xi)$ of the linear $\ol{t}$ term shows very good agreement, and is plotted in Figure~\ref{fig:linearTerm}.
We also plot the coefficient $f_2(\xi)$ of $\ol{t}^{2/3}$ in Figure~\ref{fig:t2/3term}. Here, the coefficient is slightly less numerically stable, and we cannot see the agreement quite as well.  Nonetheless, we still see fairly good agreement with the data.
%

\section{Discussion}
\label{sec:disc}

While the leading order $N^{5/3}k^{1/3}$ behavior of the free energy of GT theory was essentially known since the model was first introduced, the subleading corrections have been surprisingly difficulty to obtain analytically.  The planar resolvent was constructed in \cite{Suyama:2010hr}.  However its form did not readily lend itself to a simple expression for the free energy beyond the leading order.  Even the remarkable Fermi-gas approach to Chern-Simons-matter theories \cite{Marino:2011eh} runs into limitations when exploring higher order corrections \cite{Hong:2021bsb}.

We were able to obtain the planar free energy up to exponentially small corrections in the limit of large 't~Hooft coupling by working with the resolvent (\ref{eq:vzint}) in integral form.  The main technical observation is that the endpoints of the cuts can be obtained from $A$-cycle integrals of the resolvent integral while the derivative of the free energy can be obtained from $B$-cycle integrals.  The order of the resulting double integrals can then be swapped, leading to expressions that can be more readily worked with.  The key results are then the endpoint relations (\ref{eq:tbD}) and the free energy expression (\ref{eq:dF0dtb}).

The expressions (\ref{eq:tbD}) and (\ref{eq:dF0dtb}) in principle allow us to obtain the planar free energy $F_0(N_1,N_2,k_1,k_2)$ in the $\bar t\gg1$ limit directly in terms of the parameters of the model.  However, inverting the endpoint equations is generally non-trivial.  Nevertheless, for small differences in the Chern-Simons levels, $|k_1-k_2|\ll|k_1+k_2|$, these equations can be inverted perturbatively, assuming the self-consistent condition $|\delta|\ll1$ on the endpoints.  Focusing on the equal rank case $N_1=N_2$, or equivalently $\Delta=0$, 
we have found an explicit expansion of the free energy. If, in addition, the Chern-Simons levels are equal, we obtain the closed form expression (\ref{eq:F0equal}), which is exact up to exponentially suppressed terms.

While we have focused on the equal rank case, one can work with unequal ranks if desired.  Here some care may be needed depending on how $N_1$ and $N_2$ scale in the large-$N$ limit, as there are now two independent 't~Hooft parameters.  If the difference in ranks, $N_1-N_2$, is held fixed, then $\Delta$ is a constant, and the perturbative inversion of the endpoint equations (\ref{eq:tbD}) can be worked out as usual.  However, if $\Delta$ is not fixed, then the inversion of $\{t_1,t_2\}\leftrightarrow\{\gamma,\delta\}$ becomes more involved and the free energy as a function of two independent 't~Hooft parameters becomes less obvious.

From a technical point of view, it is possible that the way we have chosen to break the integrals into intermediate functions is not necessarily the most efficient. Many of the expressions in Appendices \ref{appendix:endpoints} and \ref{appendix:freeEnergy} are quite complicated, and one may wonder if there is a simpler parameterization that makes the formulation more elegant.  One possibility is to organize the expressions by the degree of transcendentality.  However, it is not clear if this would actually make them simpler.

One of the motivations to examine the subleading behavior of the free energy is to compare with the holographic dual.  From this point of view, it is interesting to observe that the expansion (\ref{eq:F0expand}) involves powers of  $\zeta(3)/t$.  From the supergravity point of view, this is suggestive of the $\alpha'$ expansion of the tree-level closed string effective action which starts with a term of the form $\zeta(3)\alpha'^3R^4$ \cite{Gross:1986iv}.  More generally, at higher derivative order, one expects a series of corrections of the form $\alpha'^{3(n+1)}\zeta(3)^nD^{6n}R^4$, or equivalently $\alpha'^{3(n+1)}\zeta(3)^nR^{4+3n}$, which would provide an obvious source of corrections to the dual free energy.

Of course, for now this is only a heuristic picture, as many open questions remain to be addressed before the comparison can be made rigorous.  For one thing, while the higher derivative couplings have been extensively studied for type II strings, the dual to GT theory is massive IIA supergravity, which may not receive exactly the same corrections as ordinary type II supergravity.  Nevertheless, we expect the structure to be very similar, at least if we assume a common M-theory origin.

Perhaps more importantly, advances in computing open and closed tree-level string amplitudes have provided a clearer picture of the structure of higher derivative corrections beyond $\alpha'^3R^4$.  In particular, it is known that the $\alpha'$ expansion yields terms of the form $\alpha'^{3+n}D^{2n}R^4$ (along with counterparts such as $\alpha'^{3+n}R^{4+n}$) multiplied by various combinations of $\zeta(n)$.  Assuming the free energy can be expanded only in powers of $\zeta(3)$ then demands that these other terms not proportional to $\zeta(3)^n$ do not contribute to the free energy, and hence must vanish on-shell in the gravity dual.

Finally, the form of the planar free energy, (\ref{eq:F0equal}), where the large-$t$ expansion involves a linear function of $t$ raised to a fractional power, may hint at some underlying symmetry in the $\alpha'$ expansion.  It would be interesting to study the dual massive IIA description of GT theory and to clarify some of these questions.  One obstacle in doing so is the lack of an explicit construction of the dual supergravity background beyond the limit of 
infinitesimally small Romans mass \cite{Gaiotto:2009yz}. However, we hope that such a solution may be found in the near future.

\section*{Acknowledgements}
We wish to thank J.~Hong for useful discussions.  This work was supported in part by the U.S. Department of Energy under grant DE-SC0007859.

\appendix
\section{Endpoint Computations}\label{appendix:endpoints}

After manipulating the $A$-cycle integrals for the endpoint relations, we have arrived at the expression
\begin{equation}
    t_1=\fft{\kappa_1}{2\pi^2}J_1+\fft{\kappa_2}{2\pi^2}J_2,
\label{eq:t1=J1J2}
\end{equation}
where
\begin{equation}
    J_1=\int_0^\beta dv\fft{ve^{-\fft12(\gamma+v)}I_1(v)}{2\sqrt{\cosh\fft{\alpha-v}2\sinh\fft{\beta-v}2}},
\label{eq:J1integ}
\end{equation}
with
\begin{equation}
    I_1(v)=\int_0^\beta du\,2e^{\fft12(\gamma+u)}\sqrt{\textstyle\cosh\fft{\alpha-u}2\sinh\fft{\beta-u}2}\left(\coth\fft{v-u}2+\coth\fft{v+u}2\right),
\end{equation}
up to exponentially small corrections in the large $\gamma$ limit.  (The principal value of $I_1(v)$ has to be taken in the $J_1$ integral.)  Similar expressions for $J_2$ are given in (\ref{eq:J1intdef}) and (\ref{eq:I12intdef}).  Here we carry out the integration in order to obtain the endpoint relations (\ref{eq:tbD}).

We first work on the $I_1(v)$ integral.  As it turns out, this can be integrated in closed form, with the result
\begin{align}
    I_1(v)&=2\sinh v\left(\fft\pi2-\tan^{-1}\sinh\delta\right)\nn\\
    &\qquad+2e^{\fft12(\gamma+v)}\sqrt{\textstyle\cosh\fft{\alpha-v}2\sinh\fft{\beta-v}2}\Big[-\gamma+\log(2\cosh\delta)+\log(1-e^v)\nn\\
    &\kern16em-2\log\left(\sqrt{1-e^{v-\beta}}+\sqrt{1+e^{v-\alpha}}\right)\Big]\nn\\
    &\qquad-2e^{\fft12(\gamma-v)}\sqrt{\textstyle\cosh\fft{\alpha+v}2\sinh\fft{\beta+v}2}\Big[-\gamma+\log(2\cosh\delta)+\log(1-e^{-v})\nn\\
    &\kern16em-2\log\left(\sqrt{1-e^{-v-\beta}}+\sqrt{1+e^{-v-\alpha}}\right)\Big],
\end{align}
up to exponentially small terms in the large $\gamma$ limit.  Since $I_1(v)$ is only needed for $v\in[0,\beta]$, we can further drop exponentially small terms to get
\begin{align}
    I_1(v)&=e^v\left(\fft\pi2-\tan^{-1}\sinh\delta\right)+e^\gamma\Bigl(\gamma-\log(\ft12\cosh\delta)-\log(1-e^{-v})\Bigr)\nn\\
    &\qquad+2e^{\fft12(\gamma+v)}\sqrt{\textstyle\cosh\fft{\alpha-v}2\sinh\fft{\beta-v}2}\Big[-\gamma+\log(2\cosh\delta)+\log(1-e^v)\nn\\
    &\kern16em-2\log\left(\sqrt{1-e^{v-\beta}}+\sqrt{1+e^{v-\alpha}}\right)\Big].
\label{eq:I1asymp}
\end{align}
Note that the replacement $2\sinh v\to e^v$ in the first line of this expression is not strictly valid for $v\approx0$.  However, the rest of the integrand for $J_1$ in (\ref{eq:J1integ}) is exponentially  suppressed in this limit, so there is no harm in making this substitution.

Substituting (\ref{eq:I1asymp}) into (\ref{eq:J1integ}) now gives
\begin{align}
    J_1&=2\int_0^\beta dv\Bigl[\fft{v}{\sqrt{(1+e^{v-\alpha})(1-e^{v-\beta})}}\Bigl(
    e^{v-\gamma}\left(\fft\pi2-\tan^{-1}\sinh\delta\right)\nn\\
    &\kern18em+\gamma-\log(\ft12\cosh\delta)-\log(1-e^{-v})\Bigr)\nn\\
    &\kern4em+v\left(-\gamma+\log(2\cosh\delta)+\log(e^v-1)-2\log\left(\sqrt{1-e^{v-\beta}}+\sqrt{1+e^{v+\alpha}}\right)\right)\Bigr],
\end{align}
where we flipped the sign of $1-e^v$ in the log on the second line to take the principal value into account.  Some of the integrals in the second line can be readily done.  We also integrate the final log term in the second line by parts, with the result
\begin{align}
    J_1&=-\ft13\beta^3-2\zeta(3)-\int_0^\beta dv\,v^2\left(\fft1{\sqrt{(1+e^{v-\alpha})(1-e^{v-\beta})}}-1\right)\nn\\
    &\qquad+2\int_0^\beta dv\fft{v}{\sqrt{(1+e^{v-\alpha})(1-e^{v-\beta})}}\Bigl(
    e^{v-\gamma}\left(\fft\pi2-\tan^{-1}\sinh\delta\right)\nn\\
    &\kern19em+\gamma-\log(\ft12\cosh\delta)-\log(1-e^{-v})\Bigr).
\end{align}
The final term proportional to $\log(1-e^{-v})$ is only important for $v$ close to zero.  Thus, for this term, we can replace the square root factor in the denominator by $1$ up to exponentially small terms and then integrate.  For the remaining terms, we define $x=v-\beta$ and extend the lower range of integration to $-\infty$ (which only introduces exponentially small corrections) to obtain
\begin{align}
    J_1&=-\ft13\beta^3+\beta^2\left(\gamma+\log2-\log\cosh\delta\right)\nn\\
    &\qquad+\int_{-\infty}^0 dx\left(\fft1{\sqrt{(1-e^x)(1+e^{x-2\delta})}}-1\right)\left(-(x+\beta)^2+2(x+\beta)(\gamma+\log2-\log\cosh\delta)\right)\nn\\
    &\qquad+2\int_{-\infty}^0dx\fft{(x+\beta)e^{x-\delta}}{\sqrt{(1-e^x)(1+e^{x-2\delta})}}\left(\fft\pi2-\tan^{-1}\sinh\delta\right).
\label{eq:J1expanded}
\end{align}
Recalling that $\beta=\gamma-\delta$, the first line gives the leading order $\fft23\gamma^3$ factor we expect from \eqref{eq:endptIntermed}.

To proceed, we define a set of basis integrals
\begin{align}
    f_n(\delta)&\equiv\int_{-\infty}^0 dx\,x^n\left(\fft1{\sqrt{(1-e^x)(1+e^{x-2\delta})}}-1\right),\nn\\
    g_n(\delta)&\equiv\int_{-\infty}^0dx\fft{x^ne^{x-\delta}}{\sqrt{(1-e^x)(1+e^{x-2\delta})}}.
\end{align}
Some of the integrals can be performed without too much difficulty. In particular,
\begin{equation}
    f_0(\delta)=\delta-\log(\ft12\cosh\delta),\qquad
    g_0(\delta)=\fft\pi2-\tan^{-1}\sinh\delta.
\end{equation}
With some effort it is also possible to obtain
\begin{subequations}
\begin{align}
    f_1(\delta)&=-\fft{\pi^2}3-\delta^2+2\delta\log2+\log^22+2(\cot^{-1}e^\delta)^2+(\log\cosh\delta)^2-\Li_2\left(-e^{-2\delta}\right)\nn\\
    &\qquad+\Li_2\left(\fft1{1+e^{-2\delta}}\right),\\
    g_1(\delta)&=2\cot^{-1}e^\delta(\delta+\log(\ft12\cosh\delta))-\Im\Li_2\left(\left(\fft{i+e^\delta}{-i+e^\delta}\right)^2\right).
\end{align}
\end{subequations}
This leaves just the $f_2(\delta)$ integral to be done in order to obtain a closed form result for $J_1$.  While we have not managed to analytically find an exact form for $f_2$, it can nevertheless be expanded for $\delta\ll 1$ as
\begin{align}
    f_2(\delta)=&\frac{1}{12} \left(6 \zeta (3)+4 \log ^3(2)-\pi ^2 \log 2\right)+\qty(-\frac{\pi ^2}{12}+\log ^2(2)+\pi\log 2)\delta\nn\\
    &+\frac{1}{24}((\pi -24) \pi -12 (\log2-4) \log 2)\delta ^2-\frac{1}{6} (-8+\pi  (\log 2-1)+6\log 2)\delta ^3\nn\\
    &-\frac{1}{144} \left((\pi -24) \pi +96-12 \log ^2(2)+12\log2\right)\delta ^4+\frac{1}{480}(10 (4+\pi ) \log (4)-8 (5+2 \pi ))\delta ^5\nn\\
    &+\frac{\left(2 \pi ^2-45 \pi +123-6 \log ^2(4)+15 \log4\right)}{1080}\delta ^6+\mathcal{O}\qty(\delta^7).
\end{align}

Although we have focused on $J_1$, the second integral, $J_2$, can be worked out in a similar manner, with the result
\begin{equation}
    J_2(\delta)=J_1(-\delta)+\gamma(-\pi^2-2\pi\tan^{-1}\sinh\delta)-2\pi(\delta g_0(-\delta)+g_1(-\delta)).
\end{equation}
An interesting feature of the integral (\ref{eq:J1expanded}) is that it is a precisely a cubic function of $\gamma$ up to exponentially small terms.  In particular, we find
\begin{subequations}
\begin{align}
    J_1&=\ft23\gamma^3-2\gamma^2\log(\ft12\cosh\delta)+2\gamma\left(\log^2(\ft12\cosh\delta)+(\ft12\pi-\tan^{-1}\sinh\delta)^2\right)+j_1(\delta)+\mathcal O(e^{-\gamma}),\\
    J_2&=\ft23\gamma^3-2\gamma^2\log(\ft12\cosh\delta)+2\gamma\left(\log^2(\ft12\cosh\delta)-\ft14\pi^2+(\tan^{-1}\sinh\delta)^2\right)+j_2(\delta)+\mathcal O(e^{-\gamma}),
\end{align}
\end{subequations}
where most of the complication is only in the $\gamma$-independent terms $j_1(\delta)$ and $j_2(\delta)$, defined as
\begin{subequations}
\begin{align}
     j_1(\delta)&=-\ft23\delta^3+2\delta f_1(\delta)-f_2(\delta)+2g_1(\delta)(\ft12\pi-\tan^{-1}\sinh\delta)-2\delta(\ft12\pi-\tan^{-1}\sinh\delta)^2\nn\\
     &\qquad+2(\delta^2-f_1(\delta))\log(\ft12\cosh\delta)-2\delta\log^2(\ft12\cosh\delta),\\
     j_2(\delta)&=\ft23\delta^3-2\delta f_1(-\delta)-f_2(-\delta)-2g_1(-\delta)(\ft12\pi-\tan^{-1}\sinh\delta)+2\delta(-\ft14\pi^2+(\tan^{-1}\sinh\delta)^2)\nn\\
     &\qquad+2(\delta^2-f_1(-\delta))\log(\ft12\cosh\delta)+2\delta\log^2(\ft12\cosh\delta).
\end{align}
\end{subequations}
Note that
\begin{equation}
    j_1(0)=\ft23\log^32-\ft12\pi^2\log2-\ft12\zeta(3),\qquad
    j_2(0)=\ft23\log^32+\ft12\pi^2\log2-\ft12\zeta(3).
\end{equation}
Moreover, we can see numerically that $j_2(\delta)$ is an even function, ie, $j_2(-\delta)=j_2(\delta)$. This then suggests an exact expression for the odd part of $f_2$
\begin{align}
    f_{2,o}(\delta)&=-\ft23\delta^3+2\delta f_{1,e}(\delta)-2g_{1,o}(\delta)\ft12\pi-2g_{1,e}(\delta)\tan^{-1}\sinh\delta-2\delta(-\ft14\pi^2+(\tan^{-1}\sinh\delta)^2)\nn\\
     &\qquad-2f_{1,o}(\delta)\log(\ft12\cosh\delta)-2\delta\log^2(\ft12\cosh\delta),
\end{align}
where subscripts $e$ and $o$ denote even and odd parts, respectively. This has been verified numerically. Moreover, playing around with $j_1$ and $j_2$ numerically allows one to get an empirical expression for the even part of $f_2$
\begin{align}
    f_{2,e}(\delta)=&2 \Cl_3(2\pi-\arctan\sinh\delta)+4 \delta ^2 \log (\cosh (\delta ))+\frac{\pi ^2 \delta }{6}-\frac{1}{3} 2 \log ^3(2 \text{sech}(\delta ))\nn\\
    &+2 \left(\delta ^3+\zeta (3)\right)+\log (2 \text{sech}(\delta )) \left(2 \left(\delta ^2+\log ^2(\cosh (\delta ))+\log ^2(2)\right)+\text{gd}^2(\delta )-\frac{\pi ^2}{4}\right)\nn\\
    &-\text{gd}(\delta ) \left(\pi  \delta +\Im\left(\text{Li}_2\left(\frac{\left(i+e^{\delta }\right)^2}{\left(-i+e^{\delta }\right)^2}\right)\right)-\Im\left(\text{Li}_2\left(\frac{\left(-i+e^{\delta }\right)^2}{\left(i+e^{\delta }\right)^2}\right)\right)\right)\nn\\
    &+2 \delta  \text{Li}_2\left(-e^{2 \delta }\right)+\text{Li}_2\left(\frac{1}{1+e^{2 \delta }}\right) (-\delta +\log (\text{sech}(\delta ))+\log (2))\nn\\
    &+(\delta +\log (\text{sech}(\delta ))+\log (2)) \text{Li}_2\left(\frac{1}{2} (\tanh (\delta )+1)\right),
\end{align}
where $\Cl_3$ is the Clausen function. This holds to very high precision numerically, but we have not managed to show this identity analytically.

Finally, combining $J_1$ and $J_2$ according to (\ref{eq:t1=J1J2}) gives
\begin{align}
    t_1\approx&~\fft{\kappa_1+\kappa_2}{2\pi^2}\Bigl[\ft23\gamma^3-2\gamma^2\log(\ft12\cosh\delta)+2\gamma\left(\log^2(\ft12\cosh\delta)-\ft12\pi\tan^{-1}\sinh\delta+(\tan^{-1}\sinh\delta)^2\right)\nn\\
    &+\ft12\qty(j_1(\delta)+j_2(\delta))\Bigr]+\fft{\kappa_1-\kappa_2}{2\pi^2}\Bigl[2\gamma\left(\ft14\pi^2-\ft12\pi\tan^{-1}\sinh\delta\right)+\ft12\qty(j_1(\delta)-j_2(\delta))\Bigr].
\end{align}
We can obtain a similar expression for $t_2$ by interchanging $\kappa_1\leftrightarrow\kappa_2$ and taking $\delta\to-\delta$. Taking sums and differences, and defining
\begin{equation}
    \bar t=\ft12(t_1+t_2),\qquad\Delta=\ft12(t_1-t_2),
\end{equation}
then gives the expressions (\ref{eq:tbD}) for $\bar t$ and $\Delta$ in terms of $\gamma$ and $\delta$.  Note that the odd combination $j_{2,o}$ vanishes, at least numerically.  Numerically, we also have
\begin{equation}
    j_{1,e}-j_{2,e}=\pi^2\log(\ft12\cosh\delta).
\end{equation}
However, the remaining functions $j_{1,o}$ and $j_{1,e}+j_{2,e}$ that show up in (\ref{eq:tbD}) do not seem to have similar compact expressions.

\section{Free Energy Calculations}\label{appendix:freeEnergy}

The integrals involved in evaluating the derivative of the free energy
\begin{equation}
    \fft{\partial F_0}{\partial t_1}=\fft{\kappa_1}2\beta^2-(t_1-t_2)(\Lambda-\log4)-\fft{\kappa_1}\pi K_1-\fft{\kappa_2}\pi K_2,
\end{equation}
are similar to those for determining the endpoints.  In particular, the integrals
\begin{subequations}
\begin{align}
    K_1=\int_c^ddx\fft{\log x}{\sqrt{(x-a)(x-b)(x-c)(d-x)}}I_B(x),\\ K_2=\int_a^bdx\fft{\log(-x)}{\sqrt{(x-a)(b-x)(c-x)(d-x)}}I_B(x),
\end{align}
\end{subequations}
correspond directly to the $J_1$ and $J_2$ integrals, (\ref{eq:J1J2ints}), except with $I(x)$ replaced by the $B$-cycle integral
\begin{equation}
    I_B(z)=\int_d^{e^\Lambda}\fft{dy}y\fft{\sqrt{(y-a)(y-b)(y-c)(y-d)}}{z-y}.
\end{equation}

After dropping exponentially small terms, we can write
\begin{equation}
    I_B(z)\approx\int_\beta^\Lambda du\fft{\sqrt{(e^u+e^\alpha)(e^u-e^\beta)}}{ze^{-u}-1}.
\end{equation}
This can be integrated to give
\begin{align}
    I_B(z)&=-e^\Lambda-\ft12(e^\alpha-e^\beta)\left(\Lambda+\log4+1-\log(e^\alpha+e^\beta)\right)\nn\\
    &\quad-z\left(\Lambda+\log4-\log(e^\alpha+e^\beta)\right)+2\sqrt{(e^\beta-z)(e^\alpha+z)}\tan^{-1}\sqrt{\fft{e^\alpha+z}{e^\beta-z}}.
\end{align}
Note that the first line of this expression is independent of $z$.  We can also rewrite the $K_1$ and $K_2$ integrals over the half intervals and use reflection symmetry to write
\begin{equation}
    K_1\approx\int_0^\beta dv\fft{v(I_B(e^v)-I_B(e^{-v}))}{\sqrt{(e^\alpha+e^v)(e^\beta-e^v)}},\qquad
    K_2\approx\int_0^\alpha dv\fft{v(I_B(-e^v)-I_B(-e^{-v}))}{\sqrt{(e^\alpha-e^v)(e^\beta+e^v)}},
\end{equation}
where as usual we drop exponentially small terms.  In both integrals, we only need the antisymmetric combination $I_B(z)-I_B(1/z)$.  As a result, the $z$-independent part of $I_B(z)$ drops out, and we are left with
\begin{equation}
    K_1\approx\int_0^\beta dv\fft{2v\hat I_B(e^v)}{\sqrt{(1+e^{v-\alpha})(1-e^{v-\beta})}},\qquad
    K_2\approx\int_0^\alpha dv\fft{2v\hat I_B(-e^v)}{\sqrt{(1-e^{v-\alpha})(1+e^{v-\beta})}},
\end{equation}
where
\begin{equation}
    \hat I_B(z)=e^{-\gamma}\fft{I_B(z)-I_B(z^{-1})}2.
\end{equation}

Just as with the endpoint integrals, we can work these integrals out using the explicit form of $\hat I_B(z)$.  The arctan contribution can be integrated by parts, and after some manipulation, we find
\begin{subequations}
\begin{align}
    K_1&=\beta^2(\ft12\pi-\tan^{-1}e^\delta)-(\Lambda-\gamma-\log\ft12\cosh\delta)(\beta g_0(\delta)+g_1(\delta))\nn\\
    &\qquad-\ft12(\beta^2g_0(\delta)+2\beta g_1(\delta)+g_2(\delta))-2\tan^{-1}e^\delta(\beta f_0(\delta)+f_1(\delta)),\\
    K_2&=-\alpha^2\tan^{-1}e^\delta+(\Lambda-\gamma-\log\ft12\cosh\delta)(\alpha g_0(-\delta)+g_1(-\delta))\nn\\
    &\qquad+\ft12(\alpha^2g_0(-\delta)+2\alpha g_1(-\delta)+g_2(-\delta))-2\tan^{-1}e^\delta(\alpha f_0(-\delta)+f_1(-\delta)).
\end{align}
\end{subequations}
Replacing $\alpha$ and $\beta$ with $\gamma$ and $\delta$ gives
\begin{subequations}
\begin{align}
    K_1&=(\Lambda-\gamma-\log\ft12\cosh\delta)(\gamma(-\ft12\pi+\tan^{-1}\sinh\delta)+k_1^\Lambda)+\gamma(-g_1(\delta)-2\tan^{-1}e^\delta f_0(\delta))+k_1^0,\\
    K_2&=(\Lambda-\gamma-\log\ft12\cosh\delta)(\gamma(\ft12\pi+\tan^{-1}\sinh\delta)+k_2^\Lambda)+\gamma(g_1(-\delta)-2\tan^{-1}e^\delta f_0(-\delta))+k_2^0,
\end{align}
\end{subequations}
where
\begin{equation}
    k_1^\Lambda=\delta g_0(\delta)-g_1(\delta),\qquad k_2^\Lambda=\delta g_0(-\delta)+g_1(-\delta),
\end{equation}
and
\begin{subequations}
\begin{align}
    k_1^0&=\delta g_1(\delta)-\ft12g_2(\delta)+2\tan^{-1}e^\delta(\delta f_0(\delta)-f_1(\delta)),\\
    k_2^0&=\delta g_1(-\delta)+\ft12g_2(-\delta)-2\tan^{-1}e^\delta(\delta f_0(-\delta)+f_1(-\delta)).
\end{align}
\end{subequations}

The derivative of the free energy can then be written as
\begin{align}
    \fft{\partial F_0}{\partial t_1}&=\fft{\kappa_1}2\beta^2-(\Lambda-\log4)(t_1-t_2)\nn\\
    &\qquad+\qty(\Lambda-\gamma-\log\ft12\cosh\delta)\Bigg(\fft{\kappa_1+\kappa_2}\pi\qty(-\gamma\tan^{-1}\sinh\delta-\ft12\qty(k_1^\Lambda+k_2^\Lambda))\nn\\
    &\qquad+\fft{\kappa_1-\kappa_2}2\qty(\gamma-\ft1\pi\qty(k_1^\Lambda-k_2^\Lambda))\Bigg)\nn\\
    &\qquad-\fft{\kappa_1+\kappa_2}\pi(\gamma(-g_{1,o}-2\tan^{-1}e^\delta f_{0,e})+\ft12(k_1^0+k_2^0))\nn\\
    &\qquad-\fft{\kappa_1-\kappa_2}\pi(\gamma(-g_{1,e}-2\tan^{-1}e^\delta f_{0,o})+\ft12(k_1^0-k_2^0))
\end{align}
where we have again used the $e$ and $o$ notation to denote the even and odd components of the function.  Note that the cutoff $\Lambda$ should drop out of this expression.  Comparison with the expression for $\Delta$ in (\ref{eq:tbD}) indicates that this requires the identities
\begin{equation}
    j_{1,o}=-2\pi(\delta g_{0,e}-g_{1,o}),\qquad j_{1,e}-j_{2,e}=-2\pi(\delta g_{0,o}-g_{1,e}).
\end{equation}
along with the assumed vanishing of $j_{2,o}$.  These identities do hold numerically.  The result is then
\begin{subequations}
\begin{align}
    \fft{\partial F_0}{\partial t_1}&=\fft{\kappa_1}2(\gamma-\delta)^2+(\log4-\gamma-\log\ft12\cosh\delta)(t_1-t_2)\nn\\
    &\qquad-\fft{\kappa_1+\kappa_2}\pi(\gamma(-g_{1,o}-2\tan^{-1}e^\delta f_{0,e})+\ft12(k_1^0(\delta)+k_2^0(\delta)))\nn\\
    &\qquad-\fft{\kappa_1-\kappa_2}\pi(\gamma(-g_{1,e}-2\tan^{-1}e^\delta f_{0,o})+\ft12(k_1^0(\delta)-k_2^0(\delta))),\\
    \fft{\partial F_0}{\partial t_2}&=\fft{\kappa_1}2(\gamma+\delta)^2-(\log4-\gamma-\log\ft12\cosh\delta)(t_1-t_2)\nn\\
    &\qquad-\fft{\kappa_1+\kappa_2}\pi(\gamma(g_{1,o}-2\tan^{-1}e^{-\delta}f_{0,e})+\ft12(k_1^0(-\delta)+k_2^0(-\delta)))\nn\\
    &\qquad-\fft{\kappa_1-\kappa_2}\pi(\gamma(g_{1,e}-2\tan^{-1}e^{-\delta}f_{0,o})-\ft12(k_1^0(-\delta)-k_2^0(-\delta))).
\end{align}
\end{subequations}

We now transform from $t_1$ and $t_2$ to $\bar t$ and $\Delta$. In particular, we have
\begin{equation}
   \fft{\partial F_0}{\partial\bar t}=\fft{\partial F_0}{\partial t_1}+\fft{\partial F_0}{\partial t_2},\qquad\fft{\partial F_0}{\partial\Delta}=\fft{\partial F_0}{\partial t_1}-\fft{\partial F_0}{\partial t_2}.
\end{equation}
After some simplification, we find
\begin{subequations}
\begin{align}
    \fft{\partial F_0}{\partial\bar t}&=\fft{\kappa_1+\kappa_2}2\left((\gamma-\log\ft12\cosh\delta)^2-(\log\ft12\cosh\delta)^2-\ft2\pi(k_{1,e}^0+k_{2,e}^0)+\delta^2\right)\nn\\
    &\qquad+\fft{\kappa_1-\kappa_2}2\left(-\ft2\pi(k_{1,o}^0-k_{2,o}^0)\right),\\
    \fft{\partial F_0}{\partial\Delta}&=4\Delta(\log4-\gamma-\log\ft12\cosh\delta)\nn\\
    &\qquad+\fft{\kappa_1+\kappa_2}2\left(\ft1\pi\gamma(-2\pi\delta+4g_{1,o}-4(\tan^{-1}\sinh\delta) \log\ft12\cosh\delta)-\ft2\pi(k_{1,o}^0+k_{2,o}^0)\right)\nn\\
    &\qquad+\fft{\kappa_1-\kappa_2}2\left(\gamma^2+\ft1\pi\gamma(4g_{1,e}+4\delta\tan^{-1}\sinh\delta )-\ft2\pi(k_{1,e}^0-k_{2,e}^0)+\delta^2\right).
\end{align}
\end{subequations}

Since we have an explicit expression for $f_1(\delta)$, we should be able to verify $f_1=f_{1,e}+f_{1,o}$ where
\begin{subequations}
\begin{align}
    f_{1,e}&=\ft12\delta^2-\ft1{24}\pi^2+\ft12(\log\ft12\cosh\delta)^2+\ft12(\tan^{-1}\sinh\delta)^2,\\
    f_{1,o}&=-\ft12\pi\tan^{-1}\sinh\delta-\delta\log\ft12\cosh\delta.
\end{align}
\end{subequations}
This leads to the identities
\begin{subequations}
\begin{align}
    k_{1,e}+k_{2,e}&=
    \ft\pi2\left(\delta^2+\ft1{12}\pi^2-(\log\ft12\cosh\delta)^2-(\tan^{-1}\sinh\delta)^2\right),\\
    k_{1,o}^0-k_{2,o}^0&=\ft12\pi^2\tan^{-1}\sinh\delta,
\end{align}
\end{subequations}
which results in the simple expression for the $\bar t$ derivative of the free energy given in (\ref{eq:dF0dtb}).

\bibliographystyle{JHEP}
\bibliography{cite}

\end{document}